\documentclass[preprint,authoryear,11pt]{elsarticle}
\usepackage[latin1]{inputenc}
\usepackage[T1]{fontenc}
\usepackage{amsfonts}
\usepackage[left=25mm,top=25mm,bottom=25mm,right=25mm]{geometry}
\usepackage{dsfont,amscd}
\usepackage{graphicx}
\usepackage[cyr]{aeguill}
\usepackage[pdfdisplaydoctitle=true,
            colorlinks=true,
            urlcolor=blue,
            citecolor=blue,
            linkcolor=blue,
            pdfstartview=FitH,
            pdfpagemode=None,
            bookmarksnumbered=true]{hyperref}
\usepackage{amssymb}
\usepackage{amsmath}
\usepackage{amsthm}
\usepackage{arydshln}
\usepackage{subfigure}
\usepackage{float}

\graphicspath{{fig/}}
\theoremstyle{definition}

\theoremstyle{remark}

\newcommand{\ben}{\begin{equation*}}
\newcommand{\een}{\end{equation*}}
\newcommand{\ba}{\begin{eqnarray}}
\newcommand{\ea}{\end{eqnarray}}
\newcommand{\ban}{\begin{eqnarray*}}
\newcommand{\ean}{\end{eqnarray*}}

\hypersetup{
    pdftitle={Matrix representations for 3D strain-gradient elasticity},     pdfauthor={N. Auffray, B. Kolev and M. Petitot},     pdfsubject={MSC: 74B05, 15A72},     pdfkeywords={Anisotropy, Symmetry classes, Invariant theory},     pdflang=EN     }
\input{tcilatex}
\journal{Journal of the Mechanics and Physics of Solids}

\begin{document}

\begin{frontmatter}

\title{Matrix representations for 3D strain-gradient elasticity}

\author[MSME]{N. Auffray\corref{cor1}}
\ead{Nicolas.auffray@univ-mlv.fr}
\cortext[cor1]{Corresponding author}

\author[MSME]{H. Le Quang}
\author[MSME]{Q.C. He\corref{cor2}}
\ead{qi-chang.he@univ-mlv.fr}
\cortext[cor2]{Corresponding author}

\address[MSME]{MSME, Universit\'{e} Paris-Est, Laboratoire Mod\'{e}lisation et
Simulation Multi Echelle,MSME UMR 8208 CNRS, 5 bd Descartes, 77454
Marne-la-Vall\'{e}e, France}

\begin{abstract}
The theory of first strain gradient elasticity (SGE) is widely used to model
size and non-local effects observed in materials and structures. For a
material whose microstructure is centrosymmetric, SGE is characterized by a
sixth-order elastic tensor in addition to the classical fourth-order elastic
tensor. Even though the matrix form of the sixth-order elastic tensor is
well-known in the isotropic case, its complete matrix representations seem
to remain unavailable in the anisotropic cases. In the present paper, the
explicit matrix representations of the sixth-order elastic tensor are
derived and given for all the 3D anisotropic cases in a compact and
well-structured way. These matrix representations are necessary to the
development and application of SGE for anisotropic materials.
\end{abstract}


\end{frontmatter}








{\scriptsize \setcounter{tocdepth}{2} }


\section{Introduction}

In classical continuum mechanics (\citet{TT60,TN65}), only the first
displacement gradient is involved and all the higher order displacement
gradients are neglected in measuring the deformations of a body. This usual
kinematical framework turns out not to be rich enough to describe a variety
of important mechanical and physical phenomena. In particular, the size
effects and non-local behaviors due to the discrete nature of matter at a
sufficiently small scale, the presence of microstructural defects or the
existence of internal constraints cannot be captured by classical continuum
mechanics (see, e.g., \cite{MS07} and the references cited therein for more
details). The early development of high-order (or generalized) continuum
theories of elasticity was undertaken in the 1960s and marked with the major
contributions of \cite{Tou62,Koi64,Min64,Min65,Eri68,ME68}. For the last two
decades, the development and application of high-order continuum theories
have recently gained an impetus, owing to a growing interest in modeling and
simulating size effects and non local behaviors observed in a variety of
materials, such as polycrystalline materials, geomaterials, biomaterials and
nanostructured materials, and in small size structures (see, e.g., \cite%
{FH97,NG98,LYC+03}; \cite{DSV09,DMP11,LHH12}). \textbf{At the same time, the
development of homogenization theories (see, e.g., \cite{For98,KGB04,TJA+12}%
) makes it possible to determine higher-order moduli in terms of material
local properties and microstructure. In particular, it has been recently
evidenced (\cite{ASD03,SAD11}) that microstructures can be designed to
render higher-grade effects predominant. From the numerical point of view,
theories of generalized continua, such as strain-gradient theory, can be used
to avoid explicitly meshing coarse heterogeneous materials (\cite%
{KF98,TO08,PT12}).}

The theory of first strain-gradient elasticity (SGE) proposed by \cite{Min64}
and \cite{ME68}\ is among the most important high-order continuum theories
which have been elaborated for the last half century, and it is currently
widely used. In this theory, the infinitesimal strain tensor $\mathbf{%
\varepsilon }$\ and its gradient $\mathbf{\omega }=\nabla \mathbf{%
\varepsilon }$ are linearly related to the second-order Cauchy stress
tensor\ $\mathbf{\sigma }$\ and the third-order hyperstress tensor $\mathbf{%
\tau }$ by equation (\ref{SGE}) where a fourth-order tensor $\mathbb{C}$, a
fifth-order tensor $\mathbb{M}$\ and a sixth-order tensor $\mathbb{A}$\ are
involved and verify the index permutation symmetry properties (\ref{Sym1})
and (\ref{Sym2}). The simplest one of these three tensors, $\mathbb{C}$,
defines the conventional elastic properties of a material. The study of $%
\mathbb{C}$\ had experienced a long history (\cite{Lov44}) before a complete
understanding of $\mathbb{C}$\ was achieved quite recently. In fact, only
about 20 years ago and for the first time, \cite{HD91} explicitly posed,
rigorously formulated and treated the fundamental problem of determining all
the rotational symmetry classes that the fourth-order elastic tensor $%
\mathbb{C}$\ can possess. This problem has then received the attention of
researchers from mechanics, materials science, physics, applied mathematics
and engineering (see, e.g., \cite%
{ZB94,FV96,FV97,HZ96,Xia97,CVC01,BBS04,BBS07,MN06}).\ A comprehensive
understanding of $\mathbb{C}$\ is now available in the sense that the
correct answers to the following three fundamental questions have been
provided:\newline
\qquad (a) How many symmetry classes and which symmetry classes has $\mathbb{%
C}$ ?\newline
\qquad (b) For every given symmetry class, how many independent material
parameters has $\mathbb{C}$ ?\newline
\qquad (c) For each given symmetry class, what is the explicit matrix form
of $\mathbb{C}$\ relative to an orthonormal basis ?\newline
However, these questions remains largely open in regard to the fifth-order
tensor $\mathbb{M}$\ and sixth-order tensor $\mathbb{A}$ in the theory of
SGE. Indeed, in the 3D isotropic situation, \cite{Tou62} and \cite{Min65}
gave the general form of $\mathbb{A}$\ containing 5 independent material
parameters, and \cite{Min64} showed that $\mathbb{M}$\ must be zero. In the
2D context, \cite{ABB09} derived all anisotropic matrices of $\mathbb{A}$.
At the present time, in the 3D case, few results are known. A first result
was established by \cite{DSV09}, they constructed and studied a matrix
representation of the sixth-order tensor $\mathbb{A}$ in the isotropic
situation. And, more recently, \cite{Pap11} investigates features of the $%
\mathbb{M}$ tensor in the $\mathrm{SO(3)}$-symmetry. For that symmetry class
a coupling between the the gradient and the first gradient exists. But,
apart from these results, little is known about $\mathbb{M}$ and $\mathbb{A}$%
.

In developing and applying the theory of SGE, there is currently a real need
for posing and answering the foregoing three questions about the fifth-order
tensor $\mathbb{M}$\ and sixth-order tensor $\mathbb{A}$. We first consider
materials whose microstructure is centro-symmetric.\ In this case, $\mathbb{M%
}$\ becomes zero and investigations can be carried out only on $\mathbb{A}$.
In a companion paper dedicated to 3D SGE \cite{LAH+12}, we have proved that $%
\mathbb{A}$\ has 17 symmetry classes, identified the nature of each symmetry
class and determined the number of independent material parameters of $%
\mathbb{A}$\ belonging to a given symmetry class. Nevertheless, in the
literature and in our aforementioned work, the 3D explicit matrix forms of $%
\mathbb{A}$\ have not been furnished for the 16 anisotropic symmetry
classes. This situation prevents the theory of first SGE from being
developed and applied for anisotropic materials.

Compared with the classical fourth-order tensor $\mathbb{C}$, the
sixth-order tensor $\mathbb{A}$\ is much more complex and richer. Indeed, $%
\mathbb{A}$\ has $16$ anisotropic\ symmetry classes whereas $\mathbb{C}$\
possesses $7$ ones. In this regard, remark that, for example, the transverse
hemitropy and transverse isotropy are two distinct symmetry classes for $%
\mathbb{A}$\ but shrink into one symmetry class for $\mathbb{C}$. A similar
phenomenon also occurs for the tetrahedral and cubic symmetries. In
addition, even for the same symmetry class, the number of independent
material parameters of $\mathbb{A}$\ is much higher than that of $\mathbb{C}$%
. For instance, the cubic symmetry, $\mathbb{C}$\ contains $3$ independent
parameters while $\mathbb{A}$\ comprises\ $11$ ones.

The present work aims to solve the problem of obtaining the base of the
explicit matrix representations of $\mathbb{A}$\ for all the 17 symmetry
classes. As will be seen, the complexity and richness of $\mathbb{A}$\ make
a proper solution to this problem not straightforward at all. In fact, even
though the matrix form of $\mathbb{A}$\ relative to an orthonormal basis is
known, how to express $\mathbb{A}$ as a symmetric square matrix in a compact
and well-structured way is far from being obvious.

The paper is organized as follows. In the next section, the constitutive law
of SGE is recalled and the essential results obtained by \cite{LAH+12} about
the symmetry classes of $\mathbb{A}$ are recapitulated together with the
most important concepts involved. The main results obtained by the present
work are given in section 3. They include the explicit matrix
representations of $\mathbb{A}$\ for the 17 symmetry classes, which are
rendered very compact and well-structured by proposing an original
three-to-one subscript correspondence. Each symmetry class is associated to
a simple geometric figure, and the matrix representations of $\mathbb{A}$\
are presented in such a manner that they can be directly used without
resorting to the theory of rotational groups. In section 4, the logic of the
three-to-one subscript correspondence chosen in section 3 is explained, the
general structure of the matrix representations of $\mathbb{A} $\ is
highlighted, and some salient differences between first SGE and classical
elasticity are emphasized. In section 5, a few concluding remarks are drawn.

\section{Strain-gradient elasticity}

\label{s:MSGE}

\subsection{Constitutive law}

In the (first) strain-gradient theory of linear elasticity (see, e.g., \cite%
{Min64,ME68}), the constitutive law gives the symmetric Cauchy stress tensor 
$\mathbf{\sigma }$\ and the hyperstress tensor $\mathbf{\tau }$ in terms of
the infinitesimal strain tensor $\mathbf{\varepsilon }$\ and strain-gradient
tensor $\mathbf{\omega }=\nabla \mathbf{\varepsilon }$ through the two
linear relations:%
\begin{equation}
\begin{cases}
\sigma _{ij}=C_{ijlm}\varepsilon _{lm}+M_{ijlmn}\omega _{lmn}, \\ 
\tau _{ijk}=M_{lmijk}\varepsilon _{lm}+A_{ijklmn}\omega _{lmn}.%
\end{cases}
\label{SGE}
\end{equation}%
Above, $\sigma _{ij}$, $\varepsilon _{ij}$, $\tau _{ijk}$ and $\omega
_{ijk}=\varepsilon _{ij,k}$\ are, respectively, the matrix components of $%
\mathbf{\sigma }$, $\mathbf{\varepsilon }$,\ $\mathbf{\tau }$ and $\nabla 
\mathbf{\varepsilon }$ relative to an orthonormal basis $\{\mathbf{e}_{1},%
\mathbf{e}_{2},\mathbf{e}_{2}\}$ of a three-dimensional (3D) Euclidean
space; $C_{ijlm}$, $M_{ijlmn}$ and $A_{ijklmn}$\ are the matrix components
of the fourth-, fifth- and sixth-order elastic stiffness tensors $\mathbb{C}$%
, $\mathbb{M}$ and $\mathbb{A}$, respectively. These matrix components have
the following index permutation symmetry properties:%
\begin{equation}
C_{ijlm}=C_{jilm}=C_{lmij}\text{, \ \ \ }M_{ijklm}=M_{jiklm}=M_{ijlkm},
\label{Sym1}
\end{equation}%
\begin{equation}
A_{ijklmn}=A_{jiklmn}=A_{lmnijk}.  \label{Sym2}
\end{equation}%
In the case where the microstructure of a material exhibits centro-symmetry,
the fifth-order elastic stiffness tensor $\mathbb{M}$ of this material\ is
null, so that the constitutive law (\ref{SGE}) becomes uncoupled:%
\begin{equation}
\begin{cases}
\sigma _{ij}=C_{ijlm}\varepsilon _{lm}, \\ 
\tau _{ijk}=A_{ijklmn}\omega _{lmn}.%
\end{cases}
\label{Uncpd}
\end{equation}%
In this paper, we are essentially interested in answering the question of
what are the possible different matrix forms for $\mathbb{A}$ with the index
symmetry property (\ref{Sym2}), referred to as the strain-gradient
elasticity (SGE) tensor. The same question can be posed for $\mathbb{M}$ and
will be treated in another paper.

\subsection{Symmetry classes}

In a recent paper (\cite{LAH+12}), we have solved the problem of determining
all the symmetry classes that the sixth-order SGE tensor $\mathbb{A}$ can
have. For the purpose of the present paper, we below recall some relevant
definitions and summarize the main result obtained in \cite{LAH+12}.

Let $\mathbf{Q}$\ be an element of the 3D rotation group \textrm{SO}$(3)$.
An SGE tensor $\mathbb{A}$\ is said to be invariant under the action of $%
\mathbf{Q}$ if%
\begin{equation}
Q_{io}Q_{jp}Q_{kq}Q_{lr}Q_{ms}Q_{nt}A_{opqrst}=A_{ijklmn}.  \label{Invar}
\end{equation}%
The symmetry group of $\mathbb{A}$\ is defined as the subgroup $\mathrm{G}_{%
\mathbb{A}}$ of $\mathrm{SO}(3)$ formed of all the 3D rotation tensors
leaving $\mathbb{A}$\ invariant:%
\begin{equation}
\mathrm{G}_{\mathbb{A}}=\{\mathbf{Q}\in \mathrm{SO}(3)\left\vert
Q_{io}Q_{jp}Q_{kq}Q_{lr}Q_{ms}Q_{nt}A_{opqrst}=A_{ijklmn}\right. \}\text{.}
\label{GA}
\end{equation}%
From the physical point of view, it is meaningful to consider two SGE
tensors $\mathbb{A}$\ and $\mathbb{B}$\ as exhibiting rotational symmetry of
the same kind if their symmetry groups are conjugate in the sense that%
\begin{equation}
\text{there exists a }\mathbf{Q\in }\text{ }\mathrm{SO}(3)\text{ such that }%
\mathrm{G}_{\mathrm{B}}=\mathbf{Q}\mathrm{G}_{\mathbb{A}}\mathbf{Q}^{T}.
\label{Conj}
\end{equation}%
Thus, the (rotational) symmetry class associated to an SGE tensor $\mathbb{A}
$ can be naturally defined as the set $[\mathrm{G}_{\mathbb{A}}]$ of all the
subgroups of \textrm{SO}$(3)$ conjugate to $\mathrm{G}_{\mathbb{A}}$:%
\begin{equation}
\lbrack \mathrm{G}_{\mathbb{A}}]=\{\mathrm{G}\subseteq \mathrm{SO}%
(3)\left\vert \mathrm{G}=\mathbf{Q}\mathrm{G}_{\mathbb{A}}\mathbf{Q}^{T},%
\mathbf{Q\in }\mathrm{SO}(3)\right. \}.  \label{Clasym}
\end{equation}%
In other words, the symmetry class to which $\mathbb{A}$\ belongs
corresponds to\ its symmetry group modulo $\mathrm{SO}(3)$. In fact, this
classification leads to a partition of the set consisting of all the
subgroups of $\mathrm{SO}(3)$.

For later use, we introduce some additional notations. First, the rotation
about a vector $\mathbf{a}$ through an angle $\theta \in \lbrack 0,2\pi )$
is denoted by $\mathbf{Q}(\mathbf{a},\theta )$; in particular, the rotations 
$\mathbf{Q}(\mathbf{e}_{1}+\mathbf{e}_{2}+\mathbf{e}_{3},2\pi /3)$ and $%
\mathbf{Q}[2(\sqrt{5}-1)\mathbf{e}_{2}+(\sqrt{5}+1)\mathbf{e}_{3},2\pi /3]$
are in short noted as $\mathbf{\tilde{Q}}$ for $\mathbf{\hat{Q}}$,
respectively. In what follows, use will be made of the subsequent standard
group notations: the second-order identity tensor group $\mathbf{I}$; the 2D
rotation group $\mathrm{SO}(2)$ consisting of all rotations $\mathbf{Q}$
about a 3D vector, say $\mathbf{e}_{3}$, such that $\mathbf{Q}\mathbf{e}_{3}=%
\mathbf{e}_{3}$; the 2D orthogonal group $O(2)$ composed of all orthogonal
tensors $\mathbf{Q}$ such that $\mathbf{Q}\mathbf{e}_{3}=\pm \mathbf{e}_{3}$
for a fixed direction $\mathbf{e}_{3}$; the cyclic group $\mathrm{Z}_{r}$
with $r\geq 2$ elements generated by $\mathbf{Q}(\mathbf{e}_{3},2\pi /r)$;
the dihedral group $\mathrm{D}_{r}$ $(r\geq 2)$ with $2r$ elements generated
by $\mathbf{Q}(\mathbf{e}_{3},2\pi /r)$ and $\mathbf{Q}(\mathbf{e}_{1},\pi )$%
; the tetrahedral group $\mathcal{T}$ with 12 elements generated by $\mathrm{%
D}_{2}$ and $\mathbf{\tilde{Q}}$; the octahedral group $\mathcal{O}$
containing 24 elements generated by $\mathrm{D}_{4}$ and $\mathbf{\tilde{Q}}$%
; the icosahedral group $\mathcal{I}$ having 60 elements generated by $%
\mathrm{D}_{5}$ and $\mathbf{\hat{Q}}$. Recall that the subgroups $\mathcal{T%
}$, $\mathcal{O}$ and $\mathcal{I}$ map a tetrahedron, a cube and an
icosahedron onto themselves, respectively.

In the recent paper of \cite{LAH+12}, it is proved that the number of all
possible symmetry classes for the SGE tensors is $17$. In addition, the
number of independent matrix components of an SGE\ tensor belonging to a
given symmetry class has also been determined by \cite{LAH+12}. These
results are summarized in Table 1 for the purpose of the present work. And
all the 17 symmetry classes are graphically illustrated in Figures 1 to 17.

\begin{table}[tbp]
\begin{tabular}{|c|c|c|c|c|c|}
\hline
Name & {\scriptsize Triclinic} & {\scriptsize Monoclinic} & {\scriptsize %
Orthotropic} & {\scriptsize Chirally trigonal} & {\scriptsize Trigonal} \\ 
\hline
$\lbrack \mathrm{G}_{\mathbb{A}}]$ & $\mathbf{I}$ & $[\mathrm{Z}_{2}]$ & $[%
\mathrm{D}_{2}]$ & $[\mathrm{Z}_{3}]$ & $[\mathrm{D}_{3}]$ \\ \hline
$\#_{\mathrm{indep}}(\mathbb{A})$ & $171$ & $91$ & $51$ & $57$ & $34$ \\ 
\hline\hline
Name & {\scriptsize Chirally tetragonal} & {\scriptsize Tetragonal} & 
{\scriptsize Chirally pentagonal} & {\scriptsize Pentagonal} & {\scriptsize %
Chirally hexagonal} \\ \hline
$\lbrack \mathrm{G}_{\mathbb{A}}]$ & $[\mathrm{Z}_{4}]$ & $[\mathrm{D}_{4}]$
& $[\mathrm{Z}_{5}]$ & [$\mathrm{D}_{5}]$ & $[\mathrm{Z}_{6}]$ \\ \hline
$\#_{\mathrm{indep}}(\mathbb{A})$ & $45$ & $28$ & $35$ & $23$ & $33$ \\ 
\hline\hline
Name & {\scriptsize Hexagonal} & {\scriptsize Transversely hemitropic} & 
{\scriptsize Transversely isotropic} & {\scriptsize Tetrahedral} & 
{\scriptsize Cubic} \\ \hline
$\lbrack \mathrm{G}_{\mathbb{A}}]$ & $[\mathrm{D}_{6}]$ & $[\mathrm{SO}(2)]$
& $[\mathrm{O}(2)]$ & $[\mathcal{T}]$ & $[\mathcal{O}]$ \\ \hline
$\#_{\mathrm{indep}}(\mathbb{A})$ & $22$ & $31$ & $21$ & $17$ & $11$ \\ 
\hline\hline
Name & {\scriptsize Icosahedral} & {\scriptsize Isotropic} &  &  &  \\ \hline
$\lbrack \mathrm{G}_{\mathbb{A}}]$ & $[\mathcal{I}]$ & $\mathrm{SO}(3)$ &  & 
&  \\ \hline
$\#_{\mathrm{indep}}(\mathbb{A})$ & $6$ & $5$ &  &  &  \\ \hline
\end{tabular}%
\caption{The names, the sets of subgroups $[\mathrm{G}_{\mathbb{A}}]$ and
the numbers of independent components $\#_{\mathrm{indep}}(\mathbb{A})$ for
the 17 symmetry classes of SGE.}
\end{table}

Note that the number of all possible symmetry classes for the SGE tensors is
much higher than the one for the classical elasticity tensors, since the
former is $17$ while the latter is $8$. In the totally anisotropic case
where $[\mathrm{G}_{\mathbb{A}}]=\mathbf{I}$ and in the isotropic case where 
$[\mathrm{G}_{\mathbb{A}}]=\mathrm{SO}(3)$, the number of independent
components of $\mathbb{A}$\ is equal to $171$ and $5$, respectively. This
clearly shows the SGE theory is much more complicated than the classical
elasticity theory where the corresponding numbers are $21$ and $2$,
respectively.

\section{Matrix representations of strain-gradient elasticity: main results}

\label{Sec:MatRep}

For an SGE tensor $\mathbb{A}$ belonging to one of the $17$ symmetry classes
listed in Table 1, it is very important in various situations of theoretical
and practical interest to know the \emph{explicit matrix form} of $\mathbb{A}
$\ relative to an orthonormal basis $\{\mathbf{e}_{1},\mathbf{e}_{2},\mathbf{%
e}_{3}\}$. To solve this problem in a satisfactory way, we need: (i) firstly
to identify what are the non-zero matrix components of $\mathbb{A}$ and what
are the possible relations between its non-zero components; (ii) secondly to
elaborate an appropriate representation method according to which the matrix
of $\mathbb{A}$\ is well-structured and takes a simple and compact form so
as to be used easily. The solution to the first part of the problem can be
found in the paper of \cite{LAH+12}. The solution to the second part of the
problem is still lacking and will be provided in what follows.

\subsection{Orthonormal basis and matrix component ordering}

\label{ss:OrtOrd}

Let be defined the following subspace of third-order tensors%
\begin{equation}
\mathcal{S}^{3}=\{\mathbb{T}|\mathbb{T}=\displaystyle\sum%
\limits_{i,j,k=1}^{3}T_{ijk}\mathbf{e}_{i}\otimes \mathbf{e}_{j}\otimes 
\mathbf{e}_{k},\text{ \ }T_{ijk}=T_{jik}\}  \label{3DT}
\end{equation}%
which is an 18-dimensional vector space. By (\ref{Uncpd}), an SGE tensor $%
\mathbb{A}$\ is a linear symmetric transformation from $\mathcal{S}^{3}$\ to 
$\mathcal{S}^{3}$. In order to express the strain gradient $\mathbf{\omega }$%
\ or the hyperstress tensor $\mathbf{\tau }$ as a 18-dimensional vector and
write $\mathbb{A}$\ as a $18\times 18$ symmetric matrix, we introduce the
following orthonormal basis vectors:%
\begin{equation}
\mathbf{\hat{e}}_{\alpha }=\left( \frac{1-\delta _{ij}}{\sqrt{2}}+\frac{%
\delta _{ij}}{2}\right) \left( \mathbf{e}_{i}\otimes \mathbf{e}_{j}+\mathbf{e%
}_{j}\otimes \mathbf{e}_{i}\right) \otimes \mathbf{e}_{k},\quad 1\leq \alpha
\leq 18  \label{Basis}
\end{equation}%
where the summation convention for a repeated subscript does not apply.
Then, the strain gradient $\mathbf{\omega }$, the hyperstress tensor $%
\mathbf{\tau }$ and the SGT $\mathbb{A}$\ can be expressed as%
\begin{equation}
\mathbf{\omega }=\displaystyle\sum\limits_{\alpha =1}^{18}\hat{\omega}%
_{\alpha }\mathbf{\hat{e}}_{\alpha },\text{ \ \ \ }\mathbf{\tau }=%
\displaystyle\sum\limits_{\alpha =1}^{18}\hat{\tau}_{\alpha }\mathbf{\hat{e}}%
_{\alpha },\text{ \ \ \ }\mathbb{A}=\displaystyle\sum\limits_{\alpha ,\beta
=1}^{18}\hat{A}_{\alpha \beta }\mathbf{\hat{e}}_{\alpha }\otimes \mathbf{%
\hat{e}}_{\beta },  \label{wtA}
\end{equation}%
so that the second SGE relation in (\ref{Uncpd}) can be conveniently written
in the matrix form%
\begin{equation}
\hat{\tau}_{\alpha }=\hat{A}_{\alpha \beta }\hat{\omega}_{\beta}.
\label{SGER}
\end{equation}%
Note that, with the orthonormal basis (\ref{Basis}), the relationship
between the matrix components $\hat{\omega}_{\alpha }$ and $\omega _{ijk}$,
the one between $\hat{\tau}_{\alpha }$ and $\tau _{ijk}$, and the one
between $\hat{A}_{\alpha \beta }$ and $A_{ijklmn}$ are specified by%
\begin{equation}
\hat{\omega}_{\alpha }=%
\begin{cases}
\omega _{ijk}\text{ \ if \ }i=j, \\ 
\sqrt{2}\omega _{ijk}\text{ \ if \ }i\neq j;%
\end{cases}%
\text{ \ \ \ }\hat{\tau}_{\alpha }=%
\begin{cases}
\tau _{ijk}\text{ \ if \ }i=j, \\ 
\sqrt{2}\tau _{ijk}\text{ \ if \ }i\neq j;%
\end{cases}
\label{3-to-1}
\end{equation}%
\begin{equation}
\hat{A}_{\alpha \beta }=%
\begin{cases}
A_{ijklmn}\text{ \ if \ }i=j\ \text{and}\ l=m\text{,} \\ 
\sqrt{2}A_{ijklmn}\text{ \ if \ }i\neq j\ \text{and}\ l=m\ \text{or}\ i=j\ 
\text{and}\ l\neq m, \\ 
2A_{ijklmn}\text{ \ if \ }i\neq j\ \text{and}\ l\neq m.%
\end{cases}
\label{6-to-3}
\end{equation}

It remains to choose an appropriate three-to-one subscript correspondence
between $ijk$ and $\alpha $. For the SGE matrix $\hat{A}_{\alpha \beta }$\
to be well-structured and exhibit a simple compact form for a given symmetry
group $\mathrm{G}_{\mathbb{A}}$, some criteria guiding the choice of an
efficient three-to-one subscript correspondence are necessary. The criteria
adopted in this work are detailed and explained in the next section. The
final three-to-one subscript correspondence is specified in Table 2.

\begin{table}[tbp]
\begin{center}
\begin{tabular}{|c||c|c|c|c|c||c|}
\hline
$\alpha$ & $\mathbf{1}$ & $\mathbf{2}$ & $\mathbf{3}$ & $4 $ & $5 $ &  \\ 
\hline\hline
$ijk $ & $\mathbf{111} $ & $\mathbf{221} $ & $\mathbf{122} $ & $331 $ & $133$
& Privileged direction: $1$ \\ \hline\hline
$\alpha $ & $\mathbf{6} $ & $\mathbf{7} $ & $\mathbf{8} $ & $9 $ & $10$ & 
\\ \hline
$ijk $ & $\mathbf{222} $ & $\mathbf{112} $ & $\mathbf{121} $ & $332 $ & $233$
& Privileged direction: $2$ \\ \hline\hline
$\alpha $ & $11 $ & $12 $ & $13 $ & $14 $ & $15$ &  \\ \hline
$ijk $ & $333 $ & $113 $ & $131 $ & $223 $ & $232$ & Privileged direction: $%
3 $ \\ \hline\hline
$\alpha $ & $16 $ & $17 $ & $18$ &  &  &  \\ \hline
$ijk $ & $123 $ & $132 $ & $231$ &  &  & No privileged direction \\ \hline
\end{tabular}%
\end{center}
\caption{The three-to-one subscript correspondence for 2D (in boldface in
the table)and 3D strain-gradient elasticity}
\end{table}

\subsection{Rotation matrix}

With the basis defined in (\ref{Basis}) and the three-to-one subscript
correspondence detailed in Table 2, the action of a rotation tensor $\mathbf{%
Q}\in \mathrm{SO}(3)$ on an SGE tensor $\mathbb{A}$\ can be represented by a 
$18\times 18$ rotational matrix $\hat{Q}$\ in such a way that%
\begin{equation}
Q_{io}Q_{jp}Q_{kq}Q_{lr}Q_{ms}Q_{nt}A_{opqrst}=\hat{Q}_{\alpha \beta }\hat{A}%
_{\beta \gamma }\hat{Q}_{\gamma \delta }^{T}  \label{TransL}
\end{equation}%
where%
\begin{equation}
\hat{Q}_{\alpha \beta }=\frac{1}{2}(Q_{io}Q_{jp}+Q_{ip}Q_{jo})Q_{kq}
\label{RotM}
\end{equation}%
with $\alpha $ and $\beta $\ being associated to $ijk$\ and $opq$,
respectively. Thus, formula (\ref{Invar}) expressing the invariance of $%
\mathbb{A}$\ under the action of $\mathbf{Q}$ is equivalent to%
\begin{equation}
\hat{Q}\hat{A}\hat{Q}^{T}=\hat{A}  \label{MatInv}
\end{equation}%
where $\hat{Q}$\ stands for the $18\times 18$ matrix of components $\hat{Q}%
_{\alpha \beta }$\ and $\hat{A}$\ the $18\times 18$ matrix of components $%
\hat{A}_{\alpha \beta }$.

\subsection{Matrix representations for all symmetry classes}

\label{ss:MatRep}

We are now ready to give the explicit expression of the SGE matrix $\hat{A}$
for each of the 17 symmetry classes. To be expressed in a simple and compact
way, the matrix $\hat{A}$\ for every symmetry class\ is split into
sub-matrices so as to make appear elementary building blocks. The order
adopted to specify the expressions of $\hat{A}$\ for the symmetry classes $[%
\mathrm{Z}_{k}]$ and $[\mathrm{D}_{k}]$ is $k=1,$ $2,$ $4,$ $3,$ $6,$ $5$
and $\infty $. The reason for such a choice is explained in \autoref%
{ss:GenSha}.

\subsubsection{Symmetry class characterized by $\mathbf{I}$}

\begin{figure}[tbp]
\centering
\includegraphics[scale=0.7]{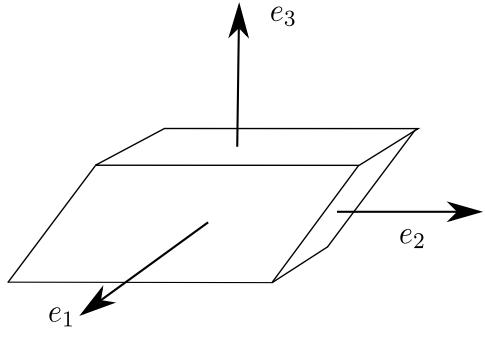}
\caption{Triclinic system ($\mathbf{I}$-invariance): the material is
completely asymmetric.}
\label{Inv}
\end{figure}

In this most general case, illustrated by figure 1, the material in question
is \textit{totally anisotropic }and the SGE matrix\ $\hat{A}$\ comprises $%
171 $ independent components. The explicit expression of $\hat{A}$\ as a
full $18\times 18$ symmetric matrix can be directly obtained by formula (\ref%
{6-to-3}). We first define

\begin{itemize}
\item the $\frac{n(n+1)}{2}$-dimensional space $\mathcal{M}^{S}(n)$
consisting of $n\times n$ symmetric matrices;

\item the $n^{2}$-dimensional space $\mathcal{M}(n)$ made of $n\times n$
matrices;

\item the $nm$-dimensional space $\mathcal{M}(n,m)$ composed of $n\times m$
matrices.
\end{itemize}

Then, we can write $\hat{A}$\ in the following way%
\begin{equation*}
A_{\mathbf{I}}=%
\begin{pmatrix}
\scriptstyle A^{(15)} & \scriptstyle B^{(25)} & \scriptstyle C^{(25)} & %
\scriptstyle D^{(15)} \\ 
& \scriptstyle E^{(15)} & \scriptstyle F^{(25)} & \scriptstyle G^{(15)} \\ 
& \scriptstyle & \scriptstyle H^{(15)} & \scriptstyle I^{(15)} \\ 
& \scriptstyle & \scriptstyle & \scriptstyle J^{(6)}%
\end{pmatrix}%
_{S}
\end{equation*}%
where the subscript $S$ indicates that the matrix is symmetric and the form
and number of independent components of each involved sub-matrix are
specified by

\begin{itemize}
\item $A^{(15)},\ E^{(15)},\ H^{(15)}\in \mathcal{M}^{S}(5)$;

\item $B^{(25)},\ C^{(25)},\ F^{(25)}\in \mathcal{M}(5)$;

\item $D^{(15)},\ G^{(15)},\ I^{(15)}\in \mathcal{M}(5,3)$;

\item $J^{(6)}\in \mathcal{M}^{S}(3)$.
\end{itemize}

For example, $A^{(15)}$\ is an element of $\mathcal{M}^{S}(5)$\ and contains
15 independent components while $B^{(25)}$\ belongs to $\mathcal{M}(5)$\ and
comprises 25 independent components.

\subsubsection{Symmetry classes $[Z_{2}]$ and $[D_{2}]$}

\begin{figure}[tbp]
\centering
\includegraphics[scale=0.7]{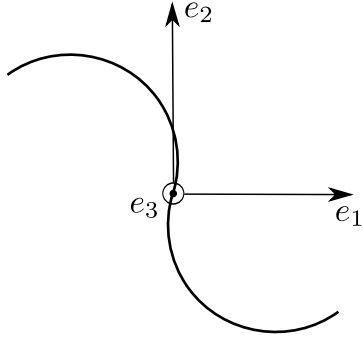}
\caption{Monoclinic system ($\mathrm{Z}_{2}$-invariance): the material is $%
\protect\pi$-invariant about $\mathbf{e}_{3}$.}
\label{Z2}
\end{figure}

\begin{figure}[tbp]
\centering
\includegraphics[scale=0.7]{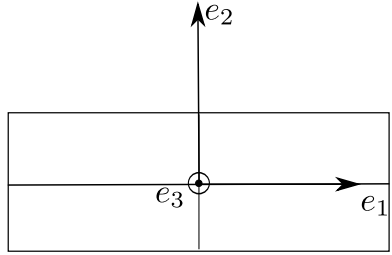}
\caption{Orthotropic system ($\mathrm{D}_{2}$-invariance): the material is $%
\protect\pi$-invariant about each of $\mathbf{e}_{1}$, $\mathbf{e}_{2}$ and $%
\mathbf{e}_{3}$.}
\label{D2}
\end{figure}


According as the symmetry class $[Z_{2}]$\ or $[D_{2}]$\ is concerned (c.f.
figures 2 and 3), the material described is \textit{monoclinic} or \textit{%
orthotropic}, and the SGE matrix\ $\hat{A}$\ contains $91$ or $51$\
independent components. Using the three-to-one subscript correspondence
given in Table 2, the SGE matrices exhibiting the $Z_{2}$-symmetry and $%
D_{2} $-symmetry are well-structured and\ take the compact forms:

\begin{equation*}
A_{\mathrm{Z}^{\mathbf{e}_{3}}_{2}}=%
\begin{pmatrix}
\scriptstyle A^{(15)} & \scriptstyle B^{(25)} & \scriptstyle0 & \scriptstyle0
\\ 
& \scriptstyle E^{(15)} & \scriptstyle 0 & \scriptstyle0 \\ 
& \scriptstyle & \scriptstyle H^{(15)} & \scriptstyle I^{(15)} \\ 
& \scriptstyle & \scriptstyle & \scriptstyle J^{(6)}%
\end{pmatrix}%
_{S}\quad ,\quad A_{\mathrm{Z}^{\mathbf{e}_{1}}_{2}}=%
\begin{pmatrix}
\scriptstyle A^{(15)} & \scriptstyle 0 & \scriptstyle0 & \scriptstyle %
D^{(15)} \\ 
& \scriptstyle E^{(15)} & \scriptstyle F^{(25)} & \scriptstyle0 \\ 
& \scriptstyle & \scriptstyle H^{(15)} & \scriptstyle 0 \\ 
& \scriptstyle & \scriptstyle & \scriptstyle J^{(6)}%
\end{pmatrix}%
_{S}
\end{equation*}%
\begin{equation*}
A_{\mathrm{D}_{2}}=%
\begin{pmatrix}
\scriptstyle A^{(15)} & \scriptstyle0 & \scriptstyle0 & \scriptstyle0 \\ 
& \scriptstyle E^{(15)} & \scriptstyle0 & \scriptstyle0 \\ 
& \scriptstyle & \scriptstyle H^{(15)} & \scriptstyle0 \\ 
& \scriptstyle & \scriptstyle & \scriptstyle J^{(6)}%
\end{pmatrix}%
_{S}
\end{equation*}%
where $A^{(15)},\ E^{(15)},\ H^{(15)}\in \mathcal{M}^{S}(5)$, $B^{(25)},
F^{(25)} \in \mathcal{M}(5)$, $I^{(15)}, D^{(15)}, \in \mathcal{M}(5,3)$\
and $J^{(6)}\in \mathcal{M}^{S}(3)$.

Here, because of its practical interest we give two conjugate
representations of the same symmetry class $[Z_{2}]$. In the first
representation the $\pi$-rotation is taken around $\mathbf{e}_{3}$ as
indicated by the notation $\mathrm{Z}^{\mathbf{e}_{3}}_{2}$, meanwhile in
the second case the rotation is taken around $\mathbf{e}_{1}$ (as indicated
by $\mathrm{Z}^{\mathbf{e}_{1}}_{2}$). The first situation is considered in
order to be coherent with the representation of the other cyclic classes, in
which the generating rotation is taken around $\mathbf{e}_{3}$. The second
representation we exhibit correspond to the common case of a monoclinic
material, the combination of the $\mathrm{Z}^{\mathbf{e}_{1}}_{2}$%
-invariance and the central inversion (always contained in the symmetry
group of any even-order tensor) leads to the existence of symmetry plane
which normal is $\mathbf{e}_{3}$.

It is remarkable that the non-zero matrix blocks of $A_{\mathrm{D}_{2}}$\
are diagonally located. Note that $A_{\mathrm{Z}^{\mathbf{e}_{1}}_{2}}$ and $%
A_{\mathrm{Z}^{\mathbf{e}_{3}}_{2}}$ are identical to $A_{\mathrm{D}_{2}}$
to within the two non-diagonal matrix blocks $B^{(25)}$ and $I^{(15)}$, in
one case and to within $E^{(25)}$ and $D^{(15)}$ in the other.

\subsubsection{Symmetry classes $[Z_{4}]$ and $[D_{4}]$}

\begin{figure}[tbp]
\centering
\includegraphics[scale=0.7]{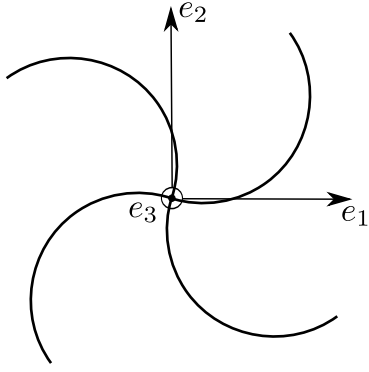}
\caption{Chirally tetragonal system ($\mathrm{Z}_{4}$-invariance): the
material is $\frac{\protect\pi}{2}$-invariant about $\mathbf{e}_{3}$.}
\label{Z4}
\end{figure}

\begin{figure}[tbp]
\centering
\includegraphics[scale=0.7]{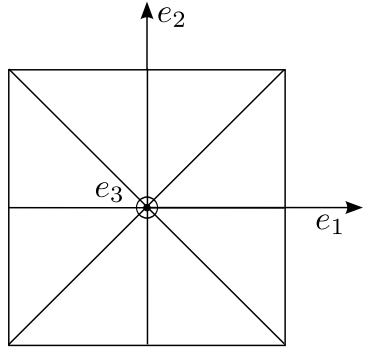}
\caption{Tetragonal system ($\mathrm{D}_{4}$-invariance): the material is $%
\frac{\protect\pi}{2}$-invariant about $\mathbf{e}_{3}$, and $\protect\pi$%
-invariant about $\left(\mathbf{Q}(\mathbf{e}_{3},k\protect\pi/4)\mathbf{e}%
_{1}\right)$ with $k\in \mathbb{Z}$.}
\label{D4}
\end{figure}


The materials characterized by the symmetry classes $[D_{4}]$\ and $[Z_{4}]$%
, shown figures 4 and 5, are said to be tetragonal and chirally tetragonal.
The numbers of independent components of $\hat{A}$\ with the $D_{4}$%
-symmetry\ and $Z_{4}$-symmetry\ are $28$ and $45$, respectively. To write
the corresponding SGE matrices in a compact way, we first introduce

\begin{itemize}
\item the $\frac{n(n-1)}{2}$-dimensional space $\mathcal{M}^{A}(n)$
consisting of $n\times n$ anti-symmetric matrices;

\item the matrices $H^{(9)}$, $I^{(7)}$\ and $J^{(4)}$\ with 9, 7 and 4\
independent components defined by
\end{itemize}

\begin{equation*}
H^{(9)}=%
\begin{pmatrix}
\scriptstyle h_{11} & \scriptstyle h_{12} & \scriptstyle h_{13} & %
\scriptstyle h_{12} & \scriptstyle h_{13} \\ 
& \scriptstyle h_{22} & \scriptstyle h_{23} & \scriptstyle h_{24} & %
\scriptstyle h_{25} \\ 
& \scriptstyle & \scriptstyle h_{33} & \scriptstyle h_{25} & \scriptstyle %
h_{35} \\ 
& \scriptstyle & \scriptstyle & \scriptstyle h_{22} & \scriptstyle h_{23} \\ 
& \scriptstyle & \scriptstyle & \scriptstyle & \scriptstyle h_{33}%
\end{pmatrix}%
_{S}\quad ,\quad I^{(7)}=%
\begin{pmatrix}
\scriptstyle0 & \scriptstyle i_{12} & \scriptstyle-i_{12} \\ 
\scriptstyle i_{21} & \scriptstyle i_{22} & \scriptstyle i_{23} \\ 
\scriptstyle i_{31} & \scriptstyle i_{32} & \scriptstyle i_{33} \\ 
\scriptstyle-i_{21} & \scriptstyle-i_{23} & \scriptstyle-i_{22} \\ 
\scriptstyle-i_{31} & \scriptstyle-i_{33} & \scriptstyle-i_{32}%
\end{pmatrix}%
\quad ,\quad J^{(4)}=%
\begin{pmatrix}
\scriptstyle j_{11} & \scriptstyle j_{12} & \scriptstyle j_{12} \\ 
& \scriptstyle j_{22} & \scriptstyle j_{23} \\ 
&  & \scriptstyle j_{22}%
\end{pmatrix}%
_{S}.
\end{equation*}%
Then, the $Z_{4}$-symmetric and $D_{4}$-symmetric SGE matrices can be
written as%
\begin{equation*}
A_{\mathrm{Z}_{4}}=%
\begin{pmatrix}
\scriptstyle A^{(15)} & \scriptstyle B^{(10)} & \scriptstyle0 & \scriptstyle0
\\ 
& \scriptstyle A^{(15)} & \scriptstyle0 & \scriptstyle0 \\ 
& \scriptstyle & \scriptstyle H^{(9)} & \scriptstyle I^{(7)} \\ 
& \scriptstyle & \scriptstyle & \scriptstyle J^{(4)}%
\end{pmatrix}%
_{S}\quad ,\quad A_{\mathrm{D}_{4}}=%
\begin{pmatrix}
\scriptstyle A^{(15)} & \scriptstyle0 & \scriptstyle0 & \scriptstyle0 \\ 
& \scriptstyle A^{(15)} & \scriptstyle0 & \scriptstyle0 \\ 
& \scriptstyle & \scriptstyle H^{(9)} & \scriptstyle0 \\ 
& \scriptstyle & \scriptstyle & \scriptstyle J^{(4)}%
\end{pmatrix}%
_{S}
\end{equation*}%
where $A^{(15)}\in \mathcal{M}^{S}(5)$ and $B^{(10)}\in \mathcal{M}^{A}(5)$%
.\ Owing to the subscript ordering specified by Table 2, both the
expressions of $A_{\mathrm{Z}_{4}}$\ and $A_{\mathrm{D}_{4}}$\ exhibit
compact structure. In addition, $A_{\mathrm{D}_{4}}$\ has a diagonal block
structure.

\subsubsection{Symmetry classes $[Z_{3}]$ and $[D_{3}]$}

\begin{figure}[tbp]
\centering
\includegraphics[scale=0.7]{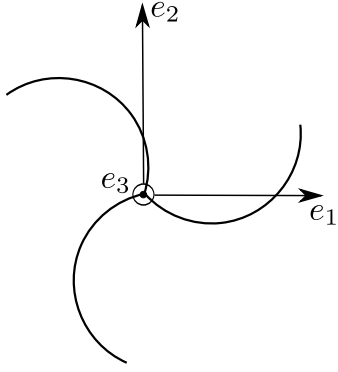}
\caption{Chirally trigonal system ($\mathrm{Z}_{3}$-invariance): the
material is $\frac{2\protect\pi}{3}$-invariant about $\mathbf{e}_{3}$.}
\label{Z3}
\end{figure}

\begin{figure}[tbp]
\centering
\includegraphics[scale=0.7]{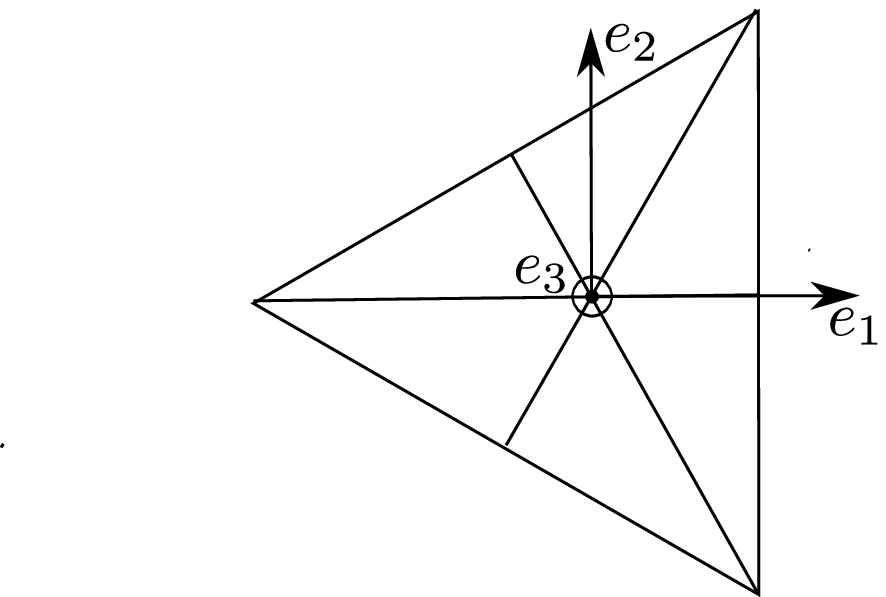}
\caption{Trigonal system ($\mathrm{D}_{3}$-invariance): the material is $%
\frac{2\protect\pi}{3}$-invariant about $\mathbf{e}_{3}$, and $\protect\pi$%
-invariant about $\left(\mathbf{Q}(\mathbf{e}_{3},k\protect\pi/3)\mathbf{e}%
_{1}\right)$, $k\in \mathbb{Z}$.}
\label{D3}
\end{figure}
%

The materials having the symmetry classes $[D_{3}]$\ and $[Z_{3}]$, shown
figures 6 and 7, are referred to as being \textit{trigonal} and \textit{%
chirally trigonal}, respectively. The numbers of independent components
contained in the corresponding matrices $A_{\mathrm{Z}_{3}}$\ and $A_{%
\mathrm{D}_{3}}$ are $57 $ and $34$, respectively. As will be seen, even if
use is made of the subscript ordering of Table 2, the matrix expressions of $%
A_{\mathrm{Z}_{3}}$\ and $A_{\mathrm{D}_{3}}$\ remain quite complex.
However, it is possible to get a good understanding of the structures of $A_{%
\mathrm{Z}_{3}}$\ and $A_{\mathrm{D}_{3}}$ by defining appropriate
independent sub-matrix blocks and making appear dependent sub-matrix blocks.
More precisely, $A_{\mathrm{Z}_{3}}$\ and $A_{\mathrm{D}_{3}}$ can be
expressed as%
\begin{equation*}
A_{\mathrm{Z}_{3}}=%
\begin{pmatrix}
\scriptstyle A^{(11)}+\eta A_{c} & \scriptstyle B^{(6)}+\theta B_{c} & %
\scriptstyle\scriptstyle C^{(3)} & \scriptstyle D^{(4)} \\ 
& \scriptstyle A^{(11)} & \scriptstyle F^{(8)} & \scriptstyle G^{(9)} \\ 
& \scriptstyle & \scriptstyle H^{(6)} & \scriptstyle I^{(4)} \\ 
& \scriptstyle & \scriptstyle & \scriptstyle J^{(4)}%
\end{pmatrix}%
_{S}+%
\begin{pmatrix}
\scriptstyle0 & \scriptstyle0 & \scriptstyle\scriptstyle f(G^{(9)}) & %
\scriptstyle f(F^{(8)}) \\ 
& \scriptstyle0 & \scriptstyle f(D^{(4)}) & \scriptstyle0 \\ 
& \scriptstyle & \scriptstyle f(J^{(4)}) & \scriptstyle0 \\ 
& \scriptstyle & \scriptstyle & \scriptstyle0%
\end{pmatrix}%
_{S},
\end{equation*}

\begin{equation*}
A_{\mathrm{D}_{3}}=%
\begin{pmatrix}
\scriptstyle A^{(11)}+\eta A_{c} & \scriptstyle0 & \scriptstyle0 & %
\scriptstyle D^{(4)} \\ 
& \scriptstyle A^{(11)} & \scriptstyle F^{(8)} & \scriptstyle0 \\ 
& \scriptstyle & \scriptstyle H^{(6)} & \scriptstyle0 \\ 
& \scriptstyle & \scriptstyle & \scriptstyle J^{(4)}%
\end{pmatrix}%
_{S}+%
\begin{pmatrix}
\scriptstyle0 & \scriptstyle0 & \scriptstyle\scriptstyle0 & \scriptstyle %
f(F^{(8)}) \\ 
& \scriptstyle0 & \scriptstyle f(D^{(4)}) & \scriptstyle0 \\ 
& \scriptstyle & \scriptstyle f(J^{(4)}) & \scriptstyle0 \\ 
& \scriptstyle & \scriptstyle & \scriptstyle0%
\end{pmatrix}%
_{S}.
\end{equation*}%
In these two expressions, $\eta $ and $\theta $ are two scalar material
parameters; $A^{(11)}$, $B^{(6)}$, $C^{(3)}$, $D^{(4)}$, $F^{(8)}$, $G^{(9)}$%
, $H^{(6)}$, $I^{(4)}$ and $J^{(4)}$ are 9 independent sub-matrices; $A_{c}$
and $B_{c}$ are two coupling matrices containing no material parameters; $%
f(G^{(9)})$, $f(F^{(8)})$, $f(D^{(4)})$ and $f(J^{(4)})$\ are the
matrix-value functions of $G^{(9)}$, $F^{(8)}$, $D^{(4)}$ and $J^{(4)}$,
respectively.\ 

First, the expression of $A^{(11)}$\ with 11 independent components is
specified by%
\begin{equation*}
A^{(11)}=%
\begin{pmatrix}
\scriptstyle a_{11} & \scriptstyle a_{12} & \scriptstyle a_{13} & %
\scriptstyle a_{14} & \scriptstyle a_{15} \\ 
& \scriptstyle a_{22} & \scriptstyle-a_{13}+\sqrt{2}\alpha _{\mathrm{III}} & %
\scriptstyle\alpha _{\mathrm{I}} & \scriptstyle\alpha _{\mathrm{II}} \\ 
&  & \scriptstyle-a_{12}+\alpha _{\mathrm{III}}^{\star } & \scriptstyle %
a_{34} & \scriptstyle a_{35} \\ 
&  &  & \scriptstyle a_{44} & \scriptstyle a_{45} \\ 
&  &  &  & \scriptstyle a_{55}%
\end{pmatrix}%
_{S}\ ,
\end{equation*}%
where%
\begin{equation*}
\alpha _{\mathrm{I}}=a_{14}-\sqrt{2}a_{34}\ ,\ \alpha _{\mathrm{II}}=a_{15}-%
\sqrt{2}a_{35}\ ,\ \alpha _{\mathrm{III}}=\frac{a_{11}-a_{22}}{2}\ ,\ \alpha
_{\mathrm{III}}^{\star }=\frac{a_{11}+a_{22}}{2}.
\end{equation*}%
Next, the expressions of $B^{(6)}$, $C^{(3)}$, $D^{(4)}$, $G^{(9)}$, $%
H^{(6)} $ and $I^{(4)}$\ are given by%
\begin{equation*}
B^{(6)}=%
\begin{pmatrix}
\scriptstyle0 & \scriptstyle b_{12} & \scriptstyle-\frac{\sqrt{2}}{2}b_{12}
& \scriptstyle b_{24}+\sqrt{2}b_{34} & \scriptstyle b_{25}+\sqrt{2}b_{35} \\ 
& \scriptstyle0 & \scriptstyle-\frac{\sqrt{2}}{2}b_{12} & \scriptstyle b_{24}
& \scriptstyle b_{25} \\ 
&  & \scriptstyle0 & \scriptstyle b_{34} & \scriptstyle b_{35} \\ 
&  &  & \scriptstyle0 & \scriptstyle b_{45} \\ 
&  &  &  & \scriptstyle0%
\end{pmatrix}%
_{A}\ ,
\end{equation*}

\begin{equation*}
C^{(3)}=%
\begin{pmatrix}
\scriptstyle c_{11} & \scriptstyle c_{12} & \scriptstyle c_{13} & %
\scriptstyle c_{12} & \scriptstyle c_{13} \\ 
-\scriptstyle c_{11} & \scriptstyle-c_{12} & \scriptstyle-c_{13} & %
\scriptstyle-c_{12} & \scriptstyle-c_{13} \\ 
\scriptstyle-\sqrt{2}c_{11} & \scriptstyle-\sqrt{2}c_{12} & \scriptstyle-%
\sqrt{2}c_{13} & \scriptstyle-\sqrt{2}c_{12} & \scriptstyle-\sqrt{2}c_{13}
\\ 
\scriptstyle0 & \scriptstyle0 & \scriptstyle0 & \scriptstyle0 & \scriptstyle0
\\ 
\scriptstyle0 & \scriptstyle0 & \scriptstyle0 & \scriptstyle0 & \scriptstyle0%
\end{pmatrix}%
\ ,\ D^{(4)}=%
\begin{pmatrix}
\scriptstyle d_{11} & \scriptstyle d_{12} & \scriptstyle-d_{12} \\ 
\scriptstyle d_{11} & \scriptstyle-d_{12} & \scriptstyle d_{12} \\ 
\scriptstyle0 & \scriptstyle-\sqrt{2}d_{12} & \scriptstyle\sqrt{2}d_{12} \\ 
\scriptstyle d_{41} & \scriptstyle0 & \scriptstyle0 \\ 
\scriptstyle d_{51} & \scriptstyle0 & \scriptstyle0%
\end{pmatrix}%
,
\end{equation*}

\begin{equation*}
F^{(8)}=%
\begin{pmatrix}
\scriptstyle f_{11} & \scriptstyle f_{12} & \scriptstyle f_{13} & %
\scriptstyle f_{14} & \scriptstyle f_{15} \\ 
\scriptstyle-f_{11} & \scriptstyle-f_{12}+\beta _{\mathrm{I}} & \scriptstyle %
f_{23} & \scriptstyle-f_{12}+\beta _{\mathrm{I}} & \scriptstyle%
-f_{15}-2\beta _{\mathrm{II}} \\ 
\scriptstyle-\sqrt{2}f_{11} & \scriptstyle-\sqrt{2}(f_{12}-\frac{3\beta _{%
\mathrm{I}}}{2}) & \scriptstyle-\sqrt{2}(f_{15}+\beta _{\mathrm{II}}) & %
\scriptstyle-\sqrt{2}(f_{12}-\frac{\beta _{\mathrm{I}}}{2}) & \scriptstyle-%
\sqrt{2}(f_{13}-\beta _{\mathrm{II}}) \\ 
\scriptstyle0 & \scriptstyle0 & \scriptstyle f_{43} & \scriptstyle0 & %
\scriptstyle-f_{43} \\ 
\scriptstyle0 & \scriptstyle0 & \scriptstyle f_{53} & \scriptstyle0 & %
\scriptstyle-f_{53} \\ 
&  &  &  & 
\end{pmatrix}%
\end{equation*}%
with%
\begin{equation*}
\beta _{\mathrm{I}}=\frac{f_{12}-f_{14}}{2}\quad ,\quad \beta _{\mathrm{II}}=%
\frac{f_{13}+f_{23}}{2},
\end{equation*}

\begin{equation*}
G^{(9)}=%
\begin{pmatrix}
\scriptstyle g_{11} & \scriptstyle g_{12} & \scriptstyle g_{13} \\ 
\scriptstyle g_{21} & \scriptstyle g_{23}-2\gamma _{\mathrm{III}} & %
\scriptstyle g_{23} \\ 
\scriptstyle\sqrt{2}\gamma _{\mathrm{I}} & \scriptstyle\sqrt{2}\gamma _{%
\mathrm{II}} & \scriptstyle\sqrt{2}(2\gamma _{\mathrm{III}}+\gamma _{\mathrm{%
II}}) \\ 
\scriptstyle g_{41} & \scriptstyle g_{42} & \scriptstyle g_{42} \\ 
\scriptstyle g_{51} & \scriptstyle g_{52} & \scriptstyle g_{52}%
\end{pmatrix}%
\end{equation*}%
with 
\begin{equation*}
\gamma _{\mathrm{I}}=\frac{g_{11}-g_{21}}{2}\quad ,\quad \gamma _{\mathrm{II}%
}=\frac{g_{13}-g_{23}}{2}\quad ,\quad \gamma _{\mathrm{III}}=\frac{%
g_{12}-g_{13}}{2},
\end{equation*}%
\begin{equation*}
H^{(6)}=%
\begin{pmatrix}
\scriptstyle h_{11} & \scriptstyle h_{12} & \scriptstyle h_{13} & %
\scriptstyle h_{12} & \scriptstyle h_{13} \\ 
& \scriptstyle h_{22} & \scriptstyle h_{23} & \scriptstyle h_{22} & %
\scriptstyle h_{23} \\ 
& \scriptstyle & \scriptstyle h_{33} & \scriptstyle h_{23} & \scriptstyle %
h_{33} \\ 
& \scriptstyle & \scriptstyle & \scriptstyle h_{22} & \scriptstyle h_{23} \\ 
& \scriptstyle & \scriptstyle & \scriptstyle & \scriptstyle h_{33}%
\end{pmatrix}%
_{S}\ ,\ I^{(4)}=%
\begin{pmatrix}
\scriptstyle0 & \scriptstyle i_{12} & \scriptstyle-i_{12} \\ 
\scriptstyle0 & \scriptstyle i_{22} & \scriptstyle-i_{22}-\sqrt{2}i_{31} \\ 
\scriptstyle i_{31} & \scriptstyle i_{32} & \scriptstyle-i_{32} \\ 
\scriptstyle0 & \scriptstyle i_{22}+\sqrt{2}i_{31} & \scriptstyle-i_{22} \\ 
\scriptstyle-i_{31} & \scriptstyle i_{32} & \scriptstyle-i_{32}%
\end{pmatrix}%
.
\end{equation*}%
The matrices $A_{c}$\ and $B_{c}$\ are independent of material parameters
and take the following forms:%
\begin{equation*}
A_{c}=%
\begin{pmatrix}
\scriptstyle1 & \scriptstyle-1 & \scriptstyle-\sqrt{2} & \scriptstyle0 & %
\scriptstyle0 \\ 
& \scriptstyle1 & \scriptstyle\sqrt{2} & \scriptstyle0 & \scriptstyle0 \\ 
& \scriptstyle & \scriptstyle2 & \scriptstyle0 & \scriptstyle0 \\ 
& \scriptstyle & \scriptstyle & \scriptstyle0 & \scriptstyle0 \\ 
& \scriptstyle & \scriptstyle & \scriptstyle & \scriptstyle0%
\end{pmatrix}%
_{S}\text{ ,}
\end{equation*}%
\begin{equation*}
B_{c}=%
\begin{pmatrix}
\scriptstyle1 & \scriptstyle0 & \scriptstyle-\frac{3\sqrt{2}}{2} & %
\scriptstyle0 & \scriptstyle0 \\ 
\scriptstyle-2 & \scriptstyle1 & \scriptstyle\frac{\sqrt{2}}{2} & %
\scriptstyle0 & \scriptstyle0 \\ 
\scriptstyle-\frac{\sqrt{2}}{2} & \scriptstyle\frac{3\sqrt{2}}{2} & %
\scriptstyle2 & \scriptstyle0 & \scriptstyle0 \\ 
\scriptstyle0 & \scriptstyle0 & \scriptstyle0 & \scriptstyle0 & \scriptstyle0
\\ 
\scriptstyle0 & \scriptstyle0 & \scriptstyle0 & \scriptstyle0 & \scriptstyle0%
\end{pmatrix}%
\text{ .}
\end{equation*}%
Finally, the matrix-value functions $f(G^{(9)})$, $f(F^{(8)})$, $f(D^{(4)})$
and $f(J^{(4)})$\ are defined by%
\begin{equation*}
f(G^{(9)})=%
\begin{pmatrix}
\scriptstyle0 & \scriptstyle-\frac{\sqrt{2}}{2}\gamma _{\mathrm{I}}^{\star }
& \scriptstyle-g_{12}-\gamma _{\mathrm{II}} & \scriptstyle\frac{\sqrt{2}}{2}%
(g_{11}+\gamma _{\mathrm{I}}) & \scriptstyle\gamma _{\mathrm{II}}^{\star }
\\ 
\scriptstyle0 & \scriptstyle-\frac{\sqrt{2}}{2}\gamma _{\mathrm{I}}^{\star }
& \scriptstyle-g_{12}+3\gamma _{\mathrm{II}}+4\gamma _{\mathrm{III}} & %
\scriptstyle\frac{\sqrt{2}}{2}(g_{11}+\gamma _{\mathrm{I}}) & \scriptstyle%
\gamma _{\mathrm{II}}^{\star } \\ 
\scriptstyle0 & \scriptstyle-2\gamma _{\mathrm{I}} & \scriptstyle0 & %
\scriptstyle0 & \scriptstyle2\sqrt{2}(\gamma _{\mathrm{III}}+\gamma _{%
\mathrm{II}}) \\ 
\scriptstyle0 & \scriptstyle-\frac{\sqrt{2}}{2}g_{41} & \scriptstyle-g_{42}
& \scriptstyle\frac{\sqrt{2}}{2}g_{41} & \scriptstyle g_{42} \\ 
\scriptstyle0 & \scriptstyle-\frac{\sqrt{2}}{2}g_{51} & \scriptstyle-g_{52}
& \scriptstyle\frac{\sqrt{2}}{2}g_{51} & \scriptstyle g_{52}%
\end{pmatrix}%
,
\end{equation*}%
\begin{equation*}
f(F^{(8)})=%
\begin{pmatrix}
\scriptstyle\sqrt{2}\alpha & \scriptstyle\beta _{\mathrm{II}} & \scriptstyle%
-2\beta _{\mathrm{III}}-\beta _{\mathrm{II}} \\ 
\scriptstyle0 & \scriptstyle\beta _{\mathrm{II}} & \scriptstyle3\beta _{%
\mathrm{II}}-2\beta _{\mathrm{III}} \\ 
\scriptstyle\alpha & \scriptstyle-2\sqrt{2}(\beta _{\mathrm{II}}-\beta _{%
\mathrm{III}}) & \scriptstyle0 \\ 
\scriptstyle0 & \scriptstyle f_{43} & \scriptstyle f_{43} \\ 
\scriptstyle0 & \scriptstyle f_{53} & \scriptstyle f_{53}%
\end{pmatrix}%
\end{equation*}%
with 
\begin{equation*}
\gamma _{\mathrm{I}}^{\star }=\frac{g_{11}+g_{21}}{2}\quad ,\quad \gamma _{%
\mathrm{II}}^{\star }=\frac{g_{13}+g_{23}}{2}\quad \text{and}\quad \beta _{%
\mathrm{III}}=\frac{f_{13}-f_{15}}{2},
\end{equation*}

\begin{equation*}
f(D^{(4)})=%
\begin{pmatrix}
\scriptstyle0 & \scriptstyle\frac{\sqrt{2}}{2}d_{11} & \scriptstyle0 & %
\scriptstyle-\frac{\sqrt{2}}{2}d_{11} & \scriptstyle0 \\ 
\scriptstyle0 & \scriptstyle\frac{\sqrt{2}}{2}d_{11} & \scriptstyle0 & %
\scriptstyle-\frac{\sqrt{2}}{2}d_{11} & \scriptstyle0 \\ 
\scriptstyle0 & \scriptstyle0 & \scriptstyle0 & \scriptstyle0 & \scriptstyle0
\\ 
\scriptstyle0 & \scriptstyle\frac{\sqrt{2}}{2}d_{41} & \scriptstyle0 & %
\scriptstyle-\frac{\sqrt{2}}{2}d_{41} & \scriptstyle0 \\ 
\scriptstyle0 & \scriptstyle\frac{\sqrt{2}}{2}d_{51} & \scriptstyle0 & %
\scriptstyle-\frac{\sqrt{2}}{2}d_{51} & \scriptstyle0%
\end{pmatrix}%
\ ,\text{ }\ f(J^{(4)})=%
\begin{pmatrix}
\scriptstyle0 & \scriptstyle0 & \scriptstyle0 & \scriptstyle0 & \scriptstyle0
\\ 
& \scriptstyle0 & \scriptstyle0 & \scriptstyle-j_{11} & \scriptstyle-\sqrt{2}%
j_{12} \\ 
& \scriptstyle & \scriptstyle0 & \scriptstyle-\sqrt{2}j_{12} & \scriptstyle%
-(j_{22}+j_{23}) \\ 
& \scriptstyle & \scriptstyle & \scriptstyle0 & \scriptstyle0 \\ 
& \scriptstyle & \scriptstyle & \scriptstyle & \scriptstyle0%
\end{pmatrix}%
_{S}
\end{equation*}

\subsubsection{Symmetry classes $[Z_{6}]$ and $[D_{6}]$}

\begin{figure}[tbp]
\centering
\includegraphics[scale=0.7]{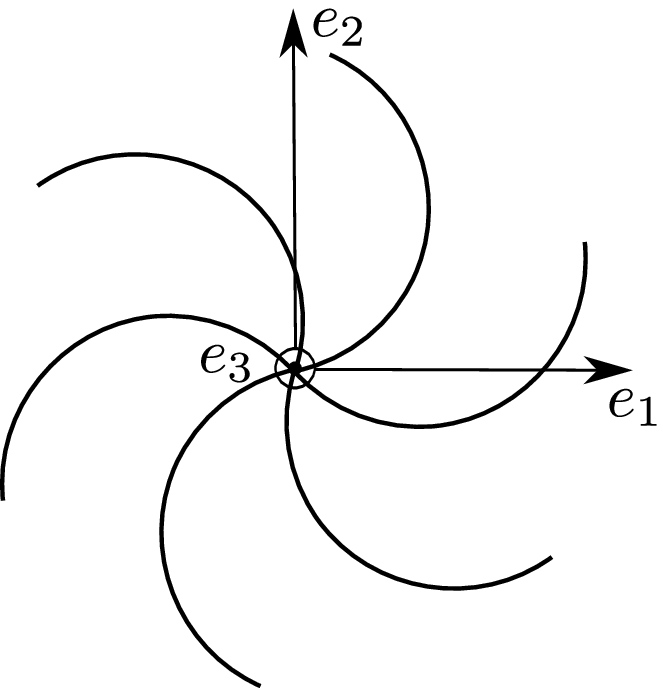}
\caption{Chirally hexagonal system ($\mathrm{Z}_{6}$-invariance): the
material is $\frac{\protect\pi}{3}$-invariant about $\mathbf{e}_{3}$}
\label{Z6}
\end{figure}

\begin{figure}[tbp]
\centering
\includegraphics[scale=0.7]{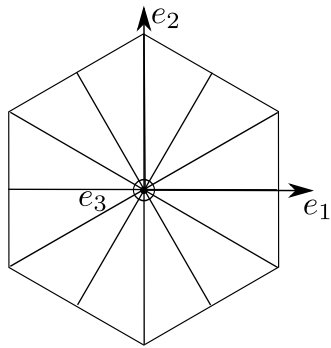}
\caption{Hexagonal system ($\mathrm{D}_{6}$-invariance): the material is $%
\frac{2\protect\pi}{6}$-invariant about $\mathbf{e}_{3}$, and $\protect\pi$%
-invariant about $\left(\mathbf{Q}(\mathbf{e}_{3},k\protect\pi/6)\mathbf{e}%
_{1}\right)$, $k\in \mathbb{Z}$.}
\label{D6}
\end{figure}
%

The \textit{hexagonal }and \textit{chirally hexagonal} materials are
described by the symmetry classes $[D_{6}]$\ and $[Z_{6}]$, illustrated by
figures 8 and 9, and have $22$ and $33$ independent parameters,
respectively. They can be considered as being degenerated from the trigonal
and chirally trigonal materials. Precisely, the associated SGE matrices $A_{%
\mathrm{Z}_{6}}$ and $A_{\mathrm{D}_{6}}$\ take the following simpler forms:

\begin{equation*}
A_{\mathrm{Z}_{6}}=%
\begin{pmatrix}
\scriptstyle A^{(11)}+\eta A_{c} & \scriptstyle B^{(6)}+\theta B_{c} & %
\scriptstyle0 & \scriptstyle0 \\ 
& \scriptstyle A^{(11)} & \scriptstyle0 & \scriptstyle0 \\ 
& \scriptstyle & \scriptstyle H^{(6)} & \scriptstyle I^{(4)} \\ 
& \scriptstyle & \scriptstyle & \scriptstyle J^{(4)}%
\end{pmatrix}%
_{S}+%
\begin{pmatrix}
\scriptstyle0 & \scriptstyle0 & \scriptstyle0 & \scriptstyle0 \\ 
& \scriptstyle0 & \scriptstyle0 & \scriptstyle0 \\ 
& \scriptstyle & \scriptstyle f(J^{(4)}) & \scriptstyle0 \\ 
& \scriptstyle & \scriptstyle & \scriptstyle0%
\end{pmatrix}%
_{S}\text{ ,}
\end{equation*}

\begin{equation*}
A_{\mathrm{D}_{6}}=%
\begin{pmatrix}
\scriptstyle A^{(11)}+\eta A_{c} & \scriptstyle0 & \scriptstyle0 & %
\scriptstyle0 \\ 
& \scriptstyle A^{(11)} & \scriptstyle0 & \scriptstyle0 \\ 
& \scriptstyle & \scriptstyle H^{(6)} & \scriptstyle0 \\ 
& \scriptstyle & \scriptstyle & \scriptstyle J^{(4)}%
\end{pmatrix}%
_{S}+%
\begin{pmatrix}
\scriptstyle0 & \scriptstyle0 & \scriptstyle0 & \scriptstyle0 \\ 
& \scriptstyle0 & \scriptstyle0 & \scriptstyle0 \\ 
& \scriptstyle & \scriptstyle f(J^{(4)}) & \scriptstyle0 \\ 
& \scriptstyle & \scriptstyle & \scriptstyle0%
\end{pmatrix}%
_{S}\text{ }
\end{equation*}%
where $\eta $ and $\theta $\ are two scalar parameters and the non-zero
matrix blocks are identical to the relevant ones given for the symmetry
class $[Z_{3}]$.

\subsubsection{Symmetry classes $[Z_{5}]$ and $[D_{5}]$.}

\begin{figure}[tbp]
\centering
\includegraphics[scale=0.7]{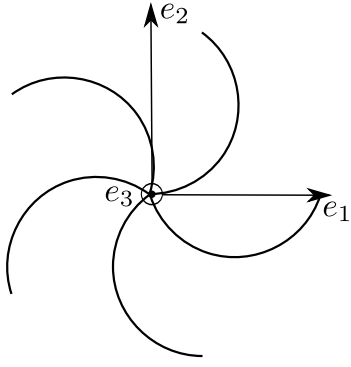}
\caption{Chirally pentagonal system ($\mathrm{Z}_{5}$-invariance): the
material is $\frac{2\protect\pi}{5}$-invariant about $\mathbf{e}_{3}$}
\label{Z5}
\end{figure}

\begin{figure}[tbp]
\centering
\includegraphics[scale=0.7]{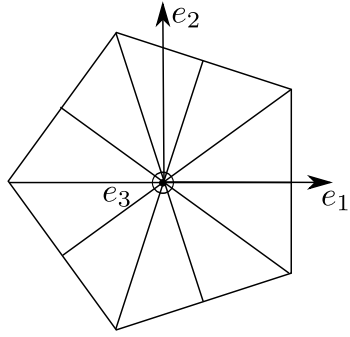}
\caption{Pentagonal system ($\mathrm{D}_{5}$-invariance): the material is $%
\frac{2\protect\pi}{5}$-invariant about $\mathbf{e}_{3}$, and $\protect\pi$%
-invariant about $\left(\mathbf{Q}(\mathbf{e}_{3},k\protect\pi/5)\mathbf{e}%
_{1}\right)$, $k\in \mathbb{Z}$.}
\label{D5}
\end{figure}


According as a material belongs to the \textit{pentagonal }symmetry class $%
[D_{5}]$\ or \textit{chirally pentagonal }symmetry $[Z_{5}]$, its number of
independent parameters is $23$ or $35$. The corresponding SGE matrices (see
figures 10 and 11) $A_{\mathrm{Z}_{5}}$\ and $A_{\mathrm{D}_{5}}$\ are given
by

\begin{equation*}
A_{\mathrm{Z}_{5}}=%
\begin{pmatrix}
\scriptstyle A^{(11)} & \scriptstyle B^{(6)} & \scriptstyle0 & \scriptstyle0
\\ 
& \scriptstyle A^{(11)} & \scriptstyle F^{(2)} & \scriptstyle G^{(2)} \\ 
& \scriptstyle & \scriptstyle H^{(6)} & \scriptstyle I^{(4)} \\ 
& \scriptstyle & \scriptstyle & \scriptstyle J^{(4)}%
\end{pmatrix}%
_{S}+%
\begin{pmatrix}
\scriptstyle0 & \scriptstyle0 & \scriptstyle\scriptstyle f(G^{(2)}) & %
\scriptstyle f(F^{(2)}) \\ 
& \scriptstyle0 & \scriptstyle0 & \scriptstyle0 \\ 
& \scriptstyle & \scriptstyle g(J^{(4)}) & \scriptstyle0 \\ 
& \scriptstyle & \scriptstyle & \scriptstyle0%
\end{pmatrix}%
_{S},
\end{equation*}

\begin{equation*}
A_{\mathrm{D}_{5}}=%
\begin{pmatrix}
\scriptstyle A^{(11)} & \scriptstyle0 & \scriptstyle0 & \scriptstyle0 \\ 
& \scriptstyle A^{(11)} & \scriptstyle F^{(2)} & \scriptstyle0 \\ 
& \scriptstyle & \scriptstyle H^{(6)} & \scriptstyle0 \\ 
& \scriptstyle & \scriptstyle & \scriptstyle J^{(4)}%
\end{pmatrix}%
_{S}+%
\begin{pmatrix}
\scriptstyle0 & \scriptstyle0 & \scriptstyle\scriptstyle0 & \scriptstyle %
f(F^{(2)}) \\ 
& \scriptstyle0 & \scriptstyle0 & \scriptstyle0 \\ 
& \scriptstyle & \scriptstyle f(J^{(4)}) & \scriptstyle0 \\ 
& \scriptstyle & \scriptstyle & \scriptstyle0%
\end{pmatrix}%
_{S}.
\end{equation*}%
\noindent In these two expressions, the sub-matrices $A^{(11)}$, $B^{(6)}$, $%
H^{(6)}$, $I^{(4)}$\ and $J^{(4)}$\ are specified in the foregoing case
where the trigonal and chirally trigonal symmetry classes, $[\mathrm{D}_{3}]$
and $[\mathrm{Z}_{3}]$,\ are concerned. The remaining sub-matrices $F^{(2)}$%
\ and $G^{(2)}$, each of which contains 2 independent components, and the
matrix-value functions $f(F^{(2)})$, $f(G^{(2)})$ and $g(J^{(4)})$\ take the
followings forms:

\begin{equation*}
F^{(2)}=%
\begin{pmatrix}
\scriptstyle0 & \scriptstyle f_{12} & \scriptstyle f_{13} & \scriptstyle%
-f_{12} & \scriptstyle-f_{13} \\ 
\scriptstyle0 & \scriptstyle-f_{12} & \scriptstyle-f_{13} & \scriptstyle %
f_{12} & \scriptstyle f_{13} \\ 
\scriptstyle0 & \scriptstyle-\sqrt{2}f_{12} & \scriptstyle-\sqrt{2}f_{13} & %
\scriptstyle\sqrt{2}f_{12} & \scriptstyle\sqrt{2}f_{13} \\ 
\scriptstyle0 & \scriptstyle0 & \scriptstyle0 & \scriptstyle0 & \scriptstyle0
\\ 
\scriptstyle0 & \scriptstyle0 & \scriptstyle0 & \scriptstyle0 & \scriptstyle0%
\end{pmatrix}%
\ ,\ \ G^{(2)}=%
\begin{pmatrix}
\scriptstyle g_{11} & \scriptstyle g_{12} & \scriptstyle g_{12} \\ 
\scriptstyle-g_{11} & \scriptstyle-g_{12} & \scriptstyle-g_{12} \\ 
\scriptstyle-\sqrt{2}g_{11} & \scriptstyle-\sqrt{2}g_{12} & \scriptstyle-%
\sqrt{2}g_{12} \\ 
\scriptstyle0 & \scriptstyle0 & \scriptstyle0 \\ 
\scriptstyle0 & \scriptstyle0 & \scriptstyle0%
\end{pmatrix}%
,
\end{equation*}

\begin{equation*}
f(F^{(2)})=%
\begin{pmatrix}
\scriptstyle-\sqrt{2}f_{12} & \scriptstyle-f_{13} & \scriptstyle-f_{13} \\ 
\scriptstyle\sqrt{2}f_{12} & \scriptstyle f_{13} & \scriptstyle f_{13} \\ 
\scriptstyle2f_{12} & \scriptstyle\sqrt{2}f_{13} & \scriptstyle\sqrt{2}f_{13}
\\ 
\scriptstyle0 & \scriptstyle0 & \scriptstyle0 \\ 
\scriptstyle0 & \scriptstyle0 & \scriptstyle0%
\end{pmatrix}%
\ ,\ \ f(G^{(2)})=%
\begin{pmatrix}
\scriptstyle0 & \scriptstyle\frac{\sqrt{2}}{2}g_{11} & \scriptstyle g_{12} & %
\scriptstyle-\frac{\sqrt{2}}{2}g_{11} & -\scriptstyle g_{12} \\ 
\scriptstyle0 & \scriptstyle-\frac{\sqrt{2}}{2}g_{11} & -\scriptstyle g_{12}
& \scriptstyle\frac{\sqrt{2}}{2}g_{11} & \scriptstyle g_{12} \\ 
\scriptstyle0 & \scriptstyle-g_{11} & \scriptstyle-\sqrt{2}g_{12} & %
\scriptstyle g_{11} & \scriptstyle\sqrt{2}g_{12} \\ 
\scriptstyle0 & \scriptstyle0 & \scriptstyle0 & \scriptstyle0 & \scriptstyle0
\\ 
\scriptstyle0 & \scriptstyle0 & \scriptstyle0 & \scriptstyle0 & \scriptstyle0%
\end{pmatrix}%
,
\end{equation*}

\begin{equation*}
g(J^{(4)})=%
\begin{pmatrix}
\scriptstyle0 & \scriptstyle0 & \scriptstyle0 & \scriptstyle0 & \scriptstyle0
\\ 
& \scriptstyle0 & \scriptstyle\sqrt{2}j_{12} & \scriptstyle-j_{11} & %
\scriptstyle0 \\ 
& \scriptstyle & \scriptstyle0 & \scriptstyle0 & \scriptstyle-(j_{22}+j_{23})
\\ 
& \scriptstyle & \scriptstyle & \scriptstyle0 & \scriptstyle\sqrt{2}j_{12}
\\ 
& \scriptstyle & \scriptstyle & \scriptstyle & \scriptstyle0%
\end{pmatrix}%
_{S}\text{ .}
\end{equation*}

\subsubsection{Symmetry classes $[\mathrm{SO}(2)]$ and $[\mathrm{O}(2)]$.}

\begin{figure}[tbp]
\centering
\includegraphics[scale=0.7]{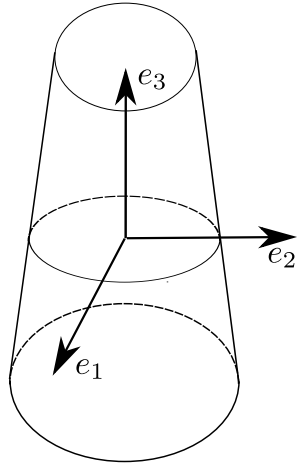}
\caption{Transversely hemitropic system ($\mathrm{SO}(2)$-invariance): the
material is $\infty$-invariant about $\mathbf{e}_{3}$}
\label{SO2}
\end{figure}

\begin{figure}[tbp]
\centering
\includegraphics[scale=0.7]{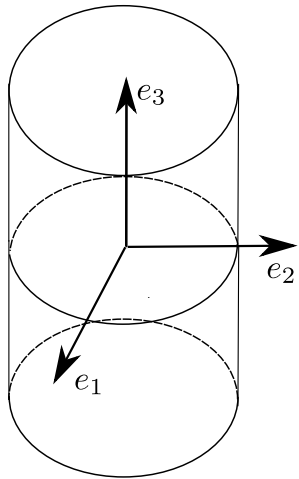}
\caption{Transversely isotropic system ($\mathrm{O}(2)$-invariance): the
material is $\infty$-invariant about $\mathbf{e}_{3}$, and $\protect\pi$%
-invariant about any in-plane axis.}
\label{O2}
\end{figure}


These two symmetry classes, shown by figures 12 and 13, characterize the 
\textit{transversely hemitropic} and \textit{transversely isotropic }%
materials, respectively. The associated SGE matrices $A_{\mathrm{SO}(2)}$\
and $A_{\mathrm{O}(2)}$, containing $31$ and $21$ independent components,
respectively, have the following expressions:

\begin{equation*}
A_{\mathrm{SO}(2)}=%
\begin{pmatrix}
\scriptstyle A^{(11)} & \scriptstyle B^{(6)} & \scriptstyle0 & \scriptstyle0
\\ 
& \scriptstyle A^{(11)} & \scriptstyle0 & \scriptstyle0 \\ 
& \scriptstyle & \scriptstyle H^{(6)} & \scriptstyle I^{(4)} \\ 
& \scriptstyle & \scriptstyle & \scriptstyle J^{(4)}%
\end{pmatrix}%
_{S}+%
\begin{pmatrix}
\scriptstyle0 & \scriptstyle0 & \scriptstyle\scriptstyle0 & \scriptstyle0 \\ 
& \scriptstyle0 & \scriptstyle0 & \scriptstyle0 \\ 
& \scriptstyle & \scriptstyle f(J^{(4)}) & \scriptstyle0 \\ 
& \scriptstyle & \scriptstyle & \scriptstyle0%
\end{pmatrix}%
_{S}\text{ ,}
\end{equation*}

\begin{equation*}
A_{\mathrm{O}(2)}=%
\begin{pmatrix}
\scriptstyle A^{(11)} & \scriptstyle0 & \scriptstyle0 & \scriptstyle0 \\ 
& \scriptstyle A^{(11)} & \scriptstyle0 & \scriptstyle0 \\ 
& \scriptstyle & \scriptstyle H^{(6)} & \scriptstyle0 \\ 
& \scriptstyle & \scriptstyle & \scriptstyle J^{(4)}%
\end{pmatrix}%
_{S}+%
\begin{pmatrix}
\scriptstyle0 & \scriptstyle0 & \scriptstyle\scriptstyle0 & \scriptstyle0 \\ 
& \scriptstyle0 & \scriptstyle0 & \scriptstyle0 \\ 
& \scriptstyle & \scriptstyle f(J^{(4)}) & \scriptstyle0 \\ 
& \scriptstyle & \scriptstyle & \scriptstyle0%
\end{pmatrix}%
_{S}\text{ ,}
\end{equation*}%
where the sub-matrices $A^{(11)}$, $B^{(6)}$, $H^{(6)}$, $I^{(4)}$\ and $%
J^{(4)}$\ are defined in studying the trigonal and chirally trigonal
symmetry classes $[\mathrm{D}_{3}]$ and $[\mathrm{Z}_{3}]$. 

\subsubsection{Symmetry classes $[\mathcal{T}]$ and $[\mathcal{O}]$.}

\begin{figure}[tbp]
\centering
\includegraphics[scale=0.7]{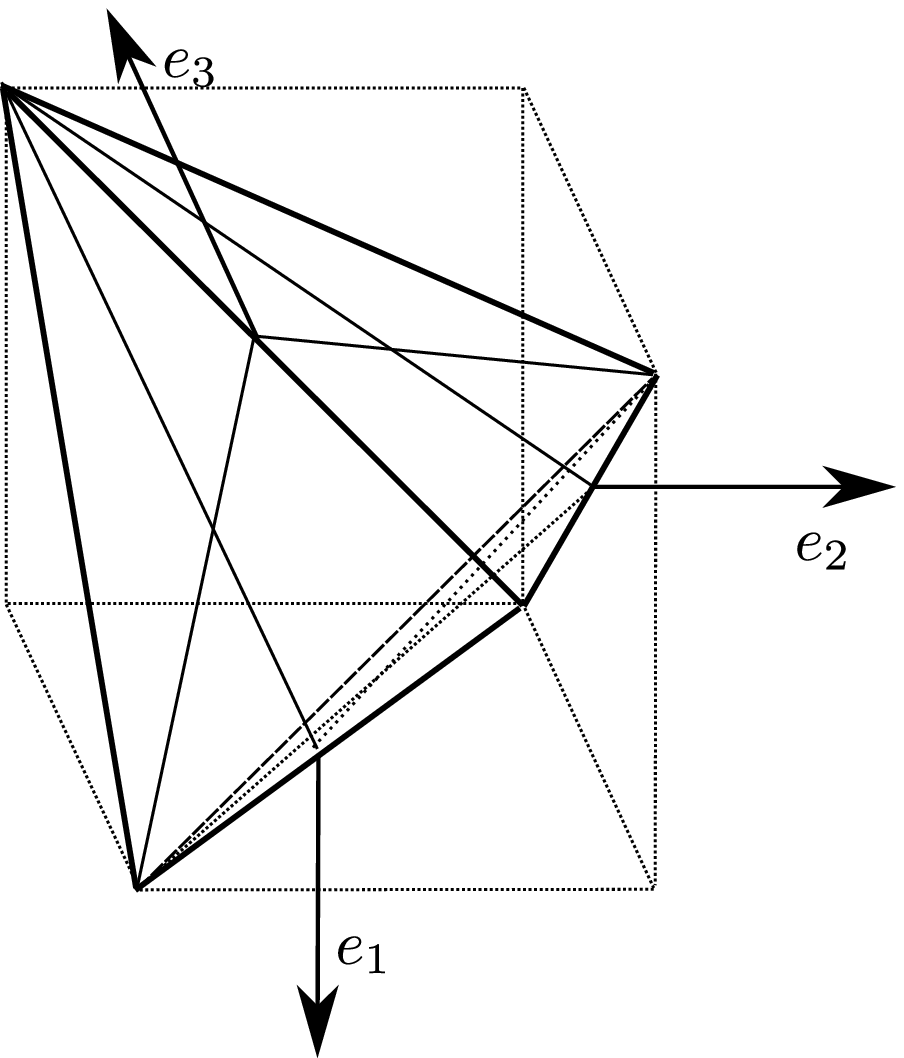}
\caption{Tetrahedral system ($\mathcal{T}$-invariance): the material is $%
\mathrm{D}_{2}$-invariant about $\mathbf{e}_{3}$ and invariant under the
permutation $(\mathbf{e}_{1}\rightarrow \mathbf{e}_{2}\rightarrow \mathbf{e}%
_{3}\rightarrow \mathbf{e}_{1})$ (The tetrahedron is drawn inside the cube
represented in fig.\protect\ref{Oct}).}
\label{Tet}
\end{figure}

\begin{figure}[tbp]
\centering
\includegraphics[scale=0.7]{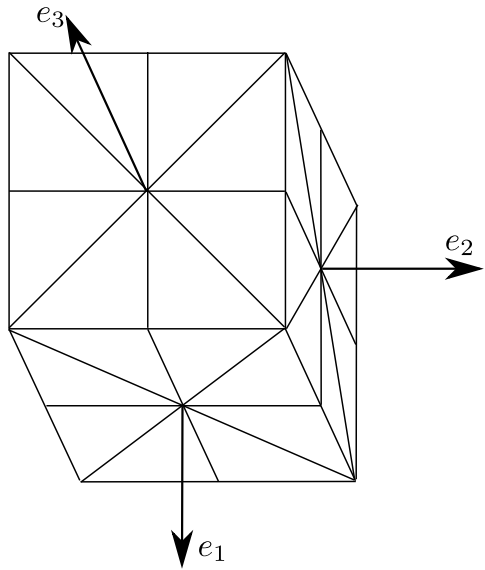}
\caption{Cubic system ($\mathcal{O}$-invariance): the material is $\mathrm{D}%
_{4}$-invariant w.r.t $\mathbf{e}_{3}$, and invariant under the permutation $%
(\mathbf{e}_{1}\rightarrow \mathbf{e}_{2}\rightarrow \mathbf{e}%
_{3}\rightarrow \mathbf{e}_{1})$.}
\label{Oct}
\end{figure}


The materials described by the symmetry classes $[\mathcal{T}]$\ and $[%
\mathcal{O}]$\, shown by figures 14 and 15, are said to be \textit{%
tetrahedral} and \textit{cubic}, respectively. The former is characterized
by 17 independent material parameters and the latter by 11 ones. The
corresponding SGE matrices $A_{\mathcal{T}}$\ and $A_{\mathcal{O}}$\ have
the expressions:

\begin{equation*}
A_{\mathcal{T}}=%
\begin{pmatrix}
\scriptstyle A^{(15)} & \scriptstyle0 & \scriptstyle0 & \scriptstyle0 \\ 
& \scriptstyle PA^{(15)}P^{T} & \scriptstyle0 & \scriptstyle0 \\ 
& \scriptstyle & \scriptstyle A^{(15)} & \scriptstyle0 \\ 
& \scriptstyle & \scriptstyle & \scriptstyle J^{(2)}%
\end{pmatrix}%
_{S}\text{ },\quad A_{\mathcal{O}}=%
\begin{pmatrix}
\scriptstyle A^{(9)} & \scriptstyle0 & \scriptstyle0 & \scriptstyle0 \\ 
& \scriptstyle A^{(9)} & \scriptstyle0 & \scriptstyle0 \\ 
& \scriptstyle & \scriptstyle A^{(9)} & \scriptstyle0 \\ 
& \scriptstyle & \scriptstyle & \scriptstyle J^{(2)}%
\end{pmatrix}%
_{S},
\end{equation*}%
where $A^{(15)}$\ is an element of $\mathcal{M}^{S}(5)$, $P$ is the
permutation matrix defined by%
\begin{equation*}
P=%
\begin{pmatrix}
1 & 0 & 0 & 0 & 0 \\ 
0 & 0 & 0 & 1 & 0 \\ 
0 & 0 & 0 & 0 & 1 \\ 
0 & 1 & 0 & 0 & 0 \\ 
0 & 0 & 1 & 0 & 0%
\end{pmatrix}%
,
\end{equation*}%
and the sub-matrices $A^{(9)}$ and $J^{(2)}$, with 9 and 2 independent
material parameters, are specified by%
\begin{equation*}
A^{(9)}=%
\begin{pmatrix}
\scriptstyle a_{11} & \scriptstyle a_{12} & \scriptstyle a_{13} & %
\scriptstyle a_{12} & \scriptstyle a_{13} \\ 
& \scriptstyle a_{22} & \scriptstyle a_{23} & \scriptstyle a_{24} & %
\scriptstyle a_{25} \\ 
& \scriptstyle & \scriptstyle a_{33} & \scriptstyle a_{25} & \scriptstyle %
a_{35} \\ 
& \scriptstyle & \scriptstyle & \scriptstyle a_{22} & \scriptstyle a_{23} \\ 
& \scriptstyle & \scriptstyle & \scriptstyle & \scriptstyle a_{33}%
\end{pmatrix}%
_{S}\ ,\ \ J^{(2)}=%
\begin{pmatrix}
\scriptstyle j_{11} & \scriptstyle j_{12} & \scriptstyle j_{12} \\ 
& \scriptstyle j_{11} & \scriptstyle j_{12} \\ 
&  & \scriptstyle j_{11}%
\end{pmatrix}%
_{S}\text{ .}
\end{equation*}%
Note that both $A_{\mathcal{T}}$\ and $A_{\mathcal{O}}$\ have a very compact
diagonal structure.

\subsubsection{Symmetry classes $[\mathcal{I}]$ and $[\mathrm{SO}(3)]$}

\begin{figure}[tbp]
\centering
\includegraphics[scale=0.6]{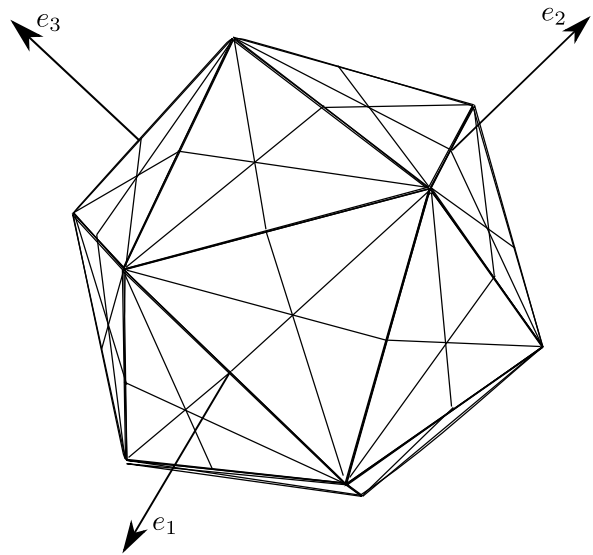}
\caption{Icosahedral ($\mathcal{I}$-invariance): the material is $\mathcal{T}
$-invariant about $\mathbf{e}_{3}$, and $\mathrm{Z}_{5}$-invariant about $%
\mathbf{v}=(\mathbf{e}_{2}+(1-\protect\phi)\mathbf{e}_{3})$. }
\label{Ico}
\end{figure}

\begin{figure}[tbp]
\centering
\includegraphics[scale=0.7]{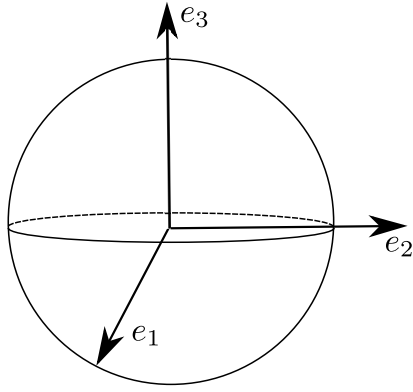}
\caption{Isotropic system ($\mathcal{O}$-invariance): the material is $%
\infty $-invariant about any axis.}
\label{Sot}
\end{figure}


The last symmetry classes $[\mathcal{I}]$\ and $[\mathrm{SO}(3)]$\, shown by
figures 15 and 16, are \textit{icosahedral} and \textit{isotropic}. The
corresponding SGE matrices $A_{\mathcal{I}}$\ and $A_{\mathrm{SO}(3)}$\
comprises 6 and 5 independent material parameters, respectively. They have
the following expressions:

\begin{equation*}
A_{\mathcal{I}}=%
\begin{pmatrix}
\scriptstyle A^{(5)}+\eta A_{\mathcal{I}}^{(c)} & \scriptstyle0 & %
\scriptstyle0 & \scriptstyle0 \\ 
& \scriptstyle P(A^{(5)}+\eta A_{\mathcal{I}}^{(c)})P^{T} & \scriptstyle0 & %
\scriptstyle0 \\ 
& \scriptstyle & \scriptstyle A^{(5)}+\eta A_{\mathcal{I}}^{(c)} & %
\scriptstyle0 \\ 
& \scriptstyle & \scriptstyle & \scriptstyle\eta J_{c}%
\end{pmatrix}%
_{S}+%
\begin{pmatrix}
\scriptstyle0 & \scriptstyle0 & \scriptstyle0 & \scriptstyle0 \\ 
& \scriptstyle0 & \scriptstyle0 & \scriptstyle0 \\ 
& \scriptstyle & \scriptstyle0 & \scriptstyle0 \\ 
& \scriptstyle & \scriptstyle & \scriptstyle f(A^{(5)})%
\end{pmatrix}%
_{S}\text{ ,}
\end{equation*}%
\begin{equation*}
A_{\mathrm{SO}(3)}=%
\begin{pmatrix}
\scriptstyle A^{(5)} & \scriptstyle0 & \scriptstyle0 & \scriptstyle0 \\ 
& \scriptstyle A^{(5)} & \scriptstyle0 & \scriptstyle0 \\ 
& \scriptstyle & \scriptstyle A^{(5)} & \scriptstyle0 \\ 
& \scriptstyle & \scriptstyle & \scriptstyle0%
\end{pmatrix}%
_{S}+%
\begin{pmatrix}
\scriptstyle0 & \scriptstyle0 & \scriptstyle0 & \scriptstyle0 \\ 
& \scriptstyle0 & \scriptstyle0 & \scriptstyle0 \\ 
& \scriptstyle & \scriptstyle0 & \scriptstyle0 \\ 
& \scriptstyle & \scriptstyle & \scriptstyle f(A^{(5)})%
\end{pmatrix}%
_{S}\text{ ,}
\end{equation*}%
where $\eta $ is a scalar material parameter, the $5\times 5$ sub-matrix $%
A^{(5)}$\ contains 5 independent material parameters, $P$ is the same
permutation matrix as the one defined in treating the tetrahedral symmetry
class $[\mathcal{T}]$, $A_{\mathcal{I}}^{(c)}$ is a $5\times 5$ sub-matrix
containing no material parameter, $J_{c}$ is a $3\times 5$ sub-matrix
involving the the golden number$\ $and $f(A^{(5)})$is a matrix-value
function of $A^{(5)}$.

More precisely, $A^{(5)}$ takes the form%
\begin{equation*}
A^{(5)}=%
\begin{pmatrix}
\scriptstyle a_{11} & \scriptstyle a_{12} & \scriptstyle a_{13} & %
\scriptstyle a_{12} & \scriptstyle a_{13} \\ 
& \scriptstyle a_{22} & \scriptstyle-a_{13}+\sqrt{2}\alpha _{\mathrm{III}} & %
\scriptstyle a_{12}-\sqrt{2}\alpha _{\mathrm{IV}}^{\star } & \scriptstyle%
\alpha _{\mathrm{IV}}^{\star } \\ 
& \scriptstyle & \scriptstyle-a_{12}+\alpha _{\mathrm{IV}} & \scriptstyle%
\alpha _{\mathrm{IV}}^{\star } & \scriptstyle a_{35} \\ 
& \scriptstyle & \scriptstyle & \scriptstyle a_{22} & \scriptstyle-a_{13}+%
\sqrt{2}\alpha _{\mathrm{III}} \\ 
& \scriptstyle & \scriptstyle & \scriptstyle & \scriptstyle-a_{12}+\alpha _{%
\mathrm{IV}}%
\end{pmatrix}%
_{S}
\end{equation*}%
with 
\begin{equation*}
\alpha _{\mathrm{III}}=\frac{a_{11}-a_{22}}{2}\ ,\ \alpha _{\mathrm{IV}%
}=a_{35}-\sqrt{2}a_{13}\ ,\ \alpha _{\mathrm{IV}}^{\star }=a_{13}-\sqrt{2}%
a_{35},
\end{equation*}%
$A_{\mathcal{I}}^{(c)}$ and $J_{c}$ are given by%
\begin{equation*}
A_{\mathcal{I}}^{(c)}=%
\begin{pmatrix}
\scriptstyle4-\phi & \scriptstyle1 & \scriptstyle2\sqrt{2} & \scriptstyle0 & %
\scriptstyle\sqrt{2} \\ 
& \scriptstyle-1 & \scriptstyle0 & \scriptstyle1-\phi & \scriptstyle0 \\ 
& \scriptstyle & \scriptstyle0 & \scriptstyle0 & \scriptstyle2-\phi \\ 
& \scriptstyle & \scriptstyle & \scriptstyle0 & \scriptstyle\sqrt{2} \\ 
& \scriptstyle & \scriptstyle & \scriptstyle & \scriptstyle2%
\end{pmatrix}%
_{S}\ ,\ J_{c}=%
\begin{pmatrix}
-1 & \overline{\phi } & \overline{\phi } \\ 
& -1 & \overline{\phi } \\ 
&  & -1%
\end{pmatrix}%
_{S}\text{ ,}
\end{equation*}%
with $\overline{\phi }$ being the conjugate of the golden number $\phi $
defined as $\overline{\phi }=\frac{1-\sqrt{5}}{2}=1-\phi $, and $f(A^{(5)})$
is specified by%
\begin{equation*}
f(A^{(5)})=%
\begin{pmatrix}
\scriptstyle\alpha _{\mathrm{V}}+\sqrt{2}\alpha _{\mathrm{IV}} & \scriptstyle%
\alpha _{\mathrm{III}}-\alpha _{\mathrm{IV}}^{\star } & \scriptstyle\alpha _{%
\mathrm{III}}-\alpha _{\mathrm{IV}}^{\star } \\ 
& \scriptstyle\alpha _{\mathrm{V}}+\sqrt{2}\alpha _{\mathrm{IV}} & %
\scriptstyle\alpha _{\mathrm{III}}-\alpha _{\mathrm{IV}}^{\star } \\ 
&  & \scriptstyle\alpha _{\mathrm{V}}+\sqrt{2}\alpha _{\mathrm{IV}}%
\end{pmatrix}%
_{S}
\end{equation*}%
with%
\begin{equation*}
\alpha _{\mathrm{V}}=a_{22}-a_{12}\text{.}
\end{equation*}

\section{Matrix representations of strain-gradient elasticity: some general
remarks}

\label{s:DerAnsi}

In the previous section, we have presented the main results about the matrix
representations for the 17 symmetry classes of strain-gradient elasticity.
These matrix representations have a very compact structure and exhibit some
general properties. In this section, we explain the reasons underlying the
three-to-one subscript correspondence specified in Table 2 and make some
general remarks.

\subsection{Matrix component ordering}

In \autoref{ss:OrtOrd}, starting from a 3-dimensional orthogonal basis $\{%
\mathbf{e}_{1},\mathbf{e}_{2},\mathbf{e}_{3}\}$, we have constructed an
18-dimensional orthonormal basis $\{\mathbf{\hat{e}}_{1},\mathbf{\hat{e}}%
_{2},...,\mathbf{\hat{e}}_{18}\}$ for $\mathcal{S}^{3}$. An SGE tensor $%
\mathbb{A}$ is a symmetric linear transformation from $\mathcal{S}^{3}$\ to $%
\mathcal{S}^{3}$. For its matrix representation relative to the basis $\{%
\mathbf{\hat{e}}_{1},\mathbf{\hat{e}}_{2},...,\mathbf{\hat{e}}_{18}\}$ to be
well-structured, some criteria have to be established to make a good choice
of the three-to-one subscript correspondence between $ijk$ and $\alpha $.
The criteria we have elaborated are explained below.

First, we consider a cubic material which is characterized by the octahedral
group $\mathcal{O}$\ graphically illustrated by figure \ref{Oct}. In this
case, the three-to-one subscript correspondence between $ijk$ and $\alpha $
is required to be such that:

\begin{description}
\item[(i) ] the matrix $A_{\mathcal{O}}$ is block-diagonal;

\item[(ii) ] each diagonal block matrix of $A_{\mathcal{O}}$ contains no
zero components;

\item[(iii) ] each diagonal block matrix of $A_{\mathcal{O}}$ is invariant
under every cyclic permutation of $\mathbf{e}_{1}$, $\mathbf{e}_{2}$ and $%
\mathbf{e}_{3}$.
\end{description}

\noindent Next, we are interested in a tetragonal material described by the
tetragonal group $\mathrm{D}_{4}$ described by figure \ref{D4}. In this
situation, we require the three-to-one subscript correspondence to be such
that

\begin{description}
\item[(iv) ] the diagonal block matrices of $A_{\mathrm{D}_{4}}$ related to
the plane $\mathbf{e}_{1}-\mathbf{e}_{2}$ are invariant under the
permutation of $\mathbf{e}_{1}$ and $\mathbf{e}_{2}$.
\end{description}

The satisfaction of the foregoing requirements (i)-(iv) has the consequence
that $A_{\mathcal{O}}$ and $A_{\mathrm{D}_{4}}$ take the forms 
\begin{equation*}
A_{\mathcal{O}}=%
\begin{pmatrix}
A & 0 & 0 & 0 \\ 
& A & 0 & 0 \\ 
&  & A & 0 \\ 
&  &  & J%
\end{pmatrix}%
_{S}\text{ },\text{ \ \ \ }A_{\mathrm{D}_{4}}=%
\begin{pmatrix}
A & 0 & 0 & 0 \\ 
& A & 0 & 0 \\ 
&  & H & 0 \\ 
&  &  & J%
\end{pmatrix}%
_{S}
\end{equation*}%
with $A\in \mathcal{M}^{S}(5)$, $H\in \mathcal{M}^{S}(5)$ and $J\in \mathcal{%
M}^{S}(3)$. In fact, the requirement (i) gives the general shape of $A_{%
\mathcal{O}}$ but does not fix the number of block matrices of $A_{\mathcal{O%
}}$. The first idea, which seems "natural", is to decompose the diagonal
part of $A_{\mathcal{O}}$\ into three $6\times 6$ block matrices. However,
use of this decomposition makes appear some zero components in each block
matrix. The elimination of zero components inside each diagonal block matrix
motivates the requirement (ii) and is performed by carrying out column/row
permutations, leading to the decomposition the diagonal part of $A_{\mathcal{%
O}}$\ or $A_{\mathrm{D}_{4}}$\ into three $5\times 5$ block matrices plus
one $3\times 3$ matrix. The requirements (iii) and (iv) are destined to
order the columns and rows within each block matrix. Precisely, the
condition (iii) leads to the invariance of every block matrix under a cubic
symmetry transformation. In particular, the first three diagonal matrices
are identical.\ The condition (iv) is imposed for the first two diagonal
block matrices of $A_{\mathrm{D}_{4}}$ to be the same and for the privileged
axis defined by $\mathbf{e}_{3}$\ to be distinguished from the privileged
axes defined by $\mathbf{e}_{1}$\ and $\mathbf{e}_{2}$ which share the same
symmetry status.

The three-to-one subscript correspondence between $ijk$ and $\alpha $
specified by Table 2 is established in agreement with the foregoing
requirements (i)-(iv). In Table 2, the second, fourth and sixth rows are
schemed out by singling out one privileged direction and describe the
interactions of the remaining directions with the privileged one. The eighth
row is "mixed" in the sense that it involves all the three directions.

Note that the fourth row is deduced from the second row by the transposition 
$(12)$ while the sixth row is obtained from the second row by an anti-cyclic
permutation $(132)$. The reason for doing so instead of deducing the fourth
and sixth rows from the second row by an cyclic permutation $(123)$\ is
twofold. First, since most of the symmetry classes of SGE are planar, it
appears judicious to privilege the planar symmetry classes, i.e., $\mathrm{%
SO(2)}$, $[\mathrm{Z}_{k}]$\ and $[\mathrm{D}_{k}]$ with $k=2,$ $3,$ $4,$ $5$
and $6$, in structuring the matrix representations of SGE. Next, our 3D
matrix representations of SGE can be easily degenerated into the 2D ones by
conserving only the elements of $\hat{A}_{\alpha \beta }$ with $\alpha
,\beta =1,$ $2,$ $3,$ $6,$ $7$ and $8$ (see the up left elements in Table 2).

\subsection{Generic matrix forms}

\label{ss:GenSha} As a consequence of the foregoing three-to-one subscript
correspondence, the matrix representations of SGE have the following generic
forms for the different symmetry classes. More precisely, for the chiral
planar symmetry classes $[\mathrm{Z}_{k}]$\ with $k=2,$ $3,$ $4,$ $5$ and $6$%
, the matrix representations exhibit the generic shapes%
\begin{equation*}
A_{\mathrm{Z}_{2r+1}}=%
\begin{pmatrix}
D_{1} & E_{12} & E_{13} & C_{1} \\ 
& D_{2} & E_{23} & C_{2} \\ 
&  & D_{3} & C_{3} \\ 
&  &  & J%
\end{pmatrix}%
_{S}\text{ },\quad A_{\mathrm{Z}_{2r}}=%
\begin{pmatrix}
D_{1} & E_{12} & 0 & 0 \\ 
& D_{2} & 0 & 0 \\ 
&  & D_{3} & C_{3} \\ 
&  &  & J%
\end{pmatrix}%
_{S}\text{ .}
\end{equation*}%
Concerning the dihedral symmetry classes $[\mathrm{D}_{k}]$ with $k=2,$ $3,$ 
$4,$ $5$ and $6$, we have%
\begin{equation*}
A_{\mathrm{D}_{2r+1}}=%
\begin{pmatrix}
D_{1} & 0 & 0 & C_{1} \\ 
& D_{2} & E_{23} & 0 \\ 
&  & D_{3} & 0 \\ 
&  &  & J%
\end{pmatrix}%
_{S}\text{ ,}\quad A_{\mathrm{D}_{2r}}=%
\begin{pmatrix}
D_{1} & 0 & 0 & 0 \\ 
& D_{2} & 0 & 0 \\ 
&  & D_{3} & 0 \\ 
&  &  & J%
\end{pmatrix}%
_{S}\text{ .}
\end{equation*}%
The matrix representations for the spatial symmetry classes $[\mathcal{T}]$, 
$[\mathcal{O}]$, $[\mathcal{I}]$ and $\mathrm{SO}(3)$ take the generic form%
\begin{equation*}
A_{\mathcal{S}}=%
\begin{pmatrix}
D & 0 & 0 & 0 \\ 
& \mathcal{P}(D) & 0 & 0 \\ 
&  & D & 0 \\ 
&  &  & J%
\end{pmatrix}%
_{S}
\end{equation*}%
where $\mathcal{P}$ means a permutation of the block matrix $D$. In
particular, for the symmetry classes $[\mathcal{O}]$ and $\mathrm{SO}(3)$,
the permutation reduces to identity, so that $\mathcal{P}(D)=D$.

From the above generic matrix forms for the different symmetry classes, it
can be seen that the elementary blocks involved in the 3D matrix
representations of SGE are of four types:

\begin{description}
\item[(a) ] $D$-type block diagonal matrices belonging to $\mathcal{M}%
^{S}(5) $;

\item[(b) ] $J$-type block diagonal matrices belonging to $\mathcal{M}%
^{S}(3) $;

\item[(c) ] $E$-type block extra-diagonal matrices belonging to $\mathcal{M}%
(5)$;

\item[(d) ] $C$-type block extra-diagonal matrices belonging to $\mathcal{M}%
(5,3)$.
\end{description}

\subsection{Remarkable differences between SGE and classical elasticity}

As recalled in section 2, SGE has $17$ symmetry classes while classical
elasticity possesses $8$ symmetry classes.

$\bullet$ First, note that the non-crystallographic symmetry classes $[%
\mathrm{Z}_{5}]$, $[\mathrm{D}_{5}]$, $[\mathcal{T}]$ and $[\mathcal{I}]$\
make sense in the case of SGE\ but disappear in classical elasticity. These
new classes are meaningful for the study of quasi-crystallographic alloys.
As is well-known, most of quasi-crystallographic ordered materials exhibit
icosahedral symmetry \citep{GQK00}. In addition $[\mathrm{Z}_{5}]$ and $[%
\mathrm{D}_{5}]$ symmetry classes are related to Penrose tilling %
\citep{Pen74}, a well-know toy-model to understand properties of
quasi-crystallographic materials.

$\bullet$ Next, SGE\ is sensitive to chirality: (i) the chiral symmetry
classes $[\mathrm{Z}_{3}]$ and $[\mathrm{Z}_{4}]$\ are present in SGE but
absent in classical elasticity; (ii) the transversely hemitropic symmetry
class $[\mathrm{SO}(2)]$\ holds for SGE but not for classical elasticity.

$\bullet $ Last, the hexagonal and chirally hexagonal symmetry classes $[%
\mathrm{D}_{6}]$ and $[\mathrm{Z}_{6}]$\ are meaningful in the case of SGE
but have the same effects as the ones of $[\mathrm{O}(2)]$\ in the case of
classical elasticity. The combination of this result with the
crystallographic restriction theorem\footnote{%
In the 2D case, the crystallographic restriction theorem states that the
orders of rotational invariance compatible with periodicity are restricted
to ${1,2,3,4}$ and ${6}$.} leads to the fact that a 2D periodic medium can
neither be transversely isotropic nor transversely hemitropic. To sum up,
both for the 2D and 3D cases, \emph{every periodic material is anisotropic
for SGE}.

The aforementioned differences constitute one of the reasons for which the
3D matrix representations of SGE are much more complex and subtler than
those of classical elasticity.

\subsection{The chiral sensitivity of SGE: comments and numerical example}

As pointed out above, SGE is sensitive to chirality. Indeed, there exist two
different types of chirality for SGE, which are explained below.

\begin{itemize}
\item The first type of chirality is related to the chiral subgroups of $%
\mathrm{O}(2)$, i.e., $\mathrm{SO}(2)$-subgroups. This is an "in-plane"
chirality which couples spatial directions, and will be called $\mathcal{S}$%
-type.  In addition, the chirality of $\mathcal{S}$-type can
be encoded both by even- and odd-order tensors. The chirality of the
sixth-order SGE tensor $\mathbb{A}$\ studied in the present paper is of $%
\mathcal{S}$-type.

\item The second type of chirality is related to the chiral subgroups of $%
\mathrm{O}(3)$ (for a detailed discussion on these subgroups, see, for
example, \cite{Ste95}), and will be qualified as $\mathcal{O}$-type. In
contrast with the chirality of $\mathcal{S}$-type, the one of $\mathcal{O}$%
-type couples the first- and second-order effects, i.e. the stress depends on the strain-gradient and the hyper-stress on the strain. Moreover, it can be encoded only by odd-order tensors. The
chiral effect of the $5$th-order tensor $\mathbb{M}$ involved in the
constitutive law (\ref{SGE}) of SGE has been studied, for example, by \cite%
{Pap11} considering a particular case. It should be noted that for some symmetry classes both phenomena appear. In such a case the couplings are of $\mathcal{SO}$-type.
\end{itemize}

\begin{figure}[H]
\centering
\includegraphics[scale=0.6]{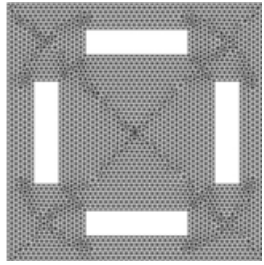}
\caption{An orthotropic cell}
\label{c_D4}
\end{figure}

\begin{figure}[H]
\centering
\includegraphics[scale=0.6]{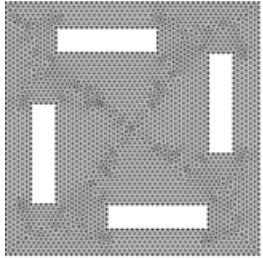}
\caption{A chirally orthotropic levogyre cell}
\label{c_Z4_L}
\end{figure}

\begin{figure}[H]
\centering
\includegraphics[scale=0.6]{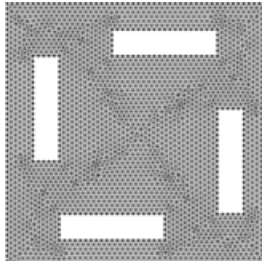}
\caption{A chirally orthotropic dextrogyre cell}
\label{c_Z4_D}
\end{figure}

To illustrate some physical implications of the chirality of $\mathcal{S}$%
-type, an example of homogenization incorporating strain gradient effects is
now numerically studied. We consider three 2D unit cells: (i) the first one
shown by Fig.\ref{c_D4} is orthotropic and belongs to the symmetry class $[%
\mathrm{D}_{4}]$; (ii) the other two ones described by Fig.\ref{c_Z4_L} and
Fig.\ref{c_Z4_D} are chirally orthotropic and falls into the symmetry class 
$[\mathrm{Z}_{4}]$.
To distinguish the  unit cell of Fig.\ref{c_Z4_L}
from that of Fig.\ref{c_Z4_D}, the former is said to be "levogyre" while
the latter is qualified as being "dextrogyre". Observe that they can be
obtained from each other by the reflection with respect to the horizontal or
vertical middle line. 
Each cell is a square of length 2 m centered at the origin, the slots are rectangles of height 0.2 m and width 0.6 m.
In the orthotropic case the the upper slot is centered at (0,0.7), and the other ones are obtained by rotation of $\frac{\pi}{2}$.
For the levogyre cell, the center of the upper slot is shifted by -0.2 m along $x_{1}$, meanwhile for the dextrogyre the shift is of 0.2 m along the same vector. The remaining slots are then obtained by symmetry operations.

The choice of a 2D example instead of a 3D one is due to the fact that a 2D
example is much simpler from the numerical standpoint but suffices for
illustrating the sensitivity of SGE to the chirality of $\mathcal{S}$-type
in a striking way. In the 2D case, the 3-to-1 correspondence to be used is
that given by the bold characters in the left up part of Table 2.

To compute the components of the sixth-order SGE tensor $\mathbb{A}$, we
apply the finite element method (FEM) to the aforementioned three unit cells
and prescribe quadratic boundary conditions (QBCs)\footnote{%
As indicated in \cite{FT11}, QBCs leads to an overestimation of $\mathbb{A}$%
. But, as shown in \cite{ABB10}, the corresponding results are qualitatively
correct.}. Such boundary conditions were proposed in \cite{GLP+97} and \cite%
{For98} and further discussed by \cite{ABB10}. The determination of the
components of $\mathbb{A}$ consists in computing the area averages of the
first moments of the stress field in a unit cell produced by elementary
QBCs. This is an extension of the classical computational homogenization
procedure for elastic heterogeneous media.

Let us first consider the orthotropic cell with 4 rectangular slots (Fig.%
\ref{c_D4}). The material forming the solid part is a linearly elastic
isotropic material with Young's modulus $E=200$ GPa and Poisson's ratio $\nu
=0.3$. Its relative density, i.e. the ratio of the area of the porous part  to the area of the solid part, is $\overline{\rho }=0.84$. Applying the FEM
and the homogenization procedure with appropriate QBCs, we obtain\footnote{%
The unit of the components is MPa.mm$^{2}$. Their values were calculated by
taking into account the relative density of the unit and rounded up.} 
\begin{equation*}
A_{\mathrm{D}_{4}}=%
\begin{pmatrix}
21320 & 8500 & -15740 & 0 & 0 & 0 \\ 
8500 & 62225 & -7720 & 0 & 0 & 0 \\ 
-15740 & -7720 & 24505 & 0 & 0 & 0 \\ 
0 & 0 & 0 & 21320 & 8500 & -15740 \\ 
0 & 0 & 0 & 8500 & 62225 & -7720 \\ 
0 & 0 & 0 & -15740 & -7720 & 24505 \\ 
&  &  &  &  & 
\end{pmatrix}%
.
\end{equation*}%
This matrix is block-diagonal and the two diagonal blocks in it are equal to
each other. Such a matrix shape is in agreement with the results presented
in $§3.3.3$.

Next, we consider the chirally orthotropic cells of Fig.\ref{c_Z4_L} and
Fig.\ref{c_Z4_D}. In these cells, the material forming the solid part is
identical to the one for the cell of Fig.\ref{c_D4} but the position of
each slot is changed. Using the same computational method as before, we
obtain%
\begin{equation*}
A_{\mathrm{Z}_{4}^{L}}=%
\begin{pmatrix}
20960 & 8150 & -14580 & 0 & 600 & -1210 \\ 
8150 & 59240 & -6560 & -600 & 0 & -2710 \\ 
-14580 & -6560 & 22350 & 1210 & 2710 & 0 \\ 
0 & -600 & 1210 & 20960 & 8150 & -14580 \\ 
600 & 0 & 2710 & 8150 & 59240 & -6560 \\ 
-1210 & -2710 & 0 & -14580 & -6560 & 22350 \\ 
&  &  &  &  & 
\end{pmatrix}%
,
\end{equation*}%
\begin{equation*}
A_{\mathrm{Z}_{4}^{D}}=%
\begin{pmatrix}
20960 & 8150 & -14580 & 0 & -600 & 1210 \\ 
8150 & 59240 & -6560 & 600 & 0 & 2710 \\ 
-14580 & -6560 & 22350 & -1210 & -2710 & 0 \\ 
0 & 600 & -1210 & 20960 & 8150 & -14580 \\ 
-600 & 0 & -2710 & 8150 & 59240 & -6560 \\ 
1210 & 2710 & 0 & -14580 & -6560 & 22350 \\ 
&  &  &  &  & 
\end{pmatrix}%
.
\end{equation*}%
In contrast with the matrix $A_{\mathrm{D}_{4}}$, the matrices $A_{\mathrm{Z}%
_{4}^{L}}$ and $A_{\mathrm{Z}_{4}^{D}}$\ are no more block-diagonal since an
antiymmetric block matrix coupling the directions 1 and 2 occurs. Further,
even though the diagonal blocks of $A_{\mathrm{Z}_{4}^{L}}$ are identical to
those of $A_{\mathrm{Z}_{4}^{D}}$, the out-of-diagonal antisymmetric blocks
of $A_{\mathrm{Z}_{4}^{L}}$ are however different from those of $A_{\mathrm{Z%
}_{4}^{D}}$ by a sign.

The comparison between the matrices $A_{\mathrm{D}_{4}}$, $A_{\mathrm{Z}%
_{4}^{L}}$ and $A_{\mathrm{Z}_{4}^{D}}$\ clearly shows the chiral
sensitivity of SGE. To get more insight, let us examine the first column of $%
\mathbb{A}$ in detail. The elements of this column are determined with the
aid of the following QBC%
\begin{equation*}
\begin{cases}
u_{1}(x_{1},x_{2})=\frac{1}{2}x_{1}^{2} \\ 
u_{2}(x_{1},x_{2})=0 \\ 
\end{cases}%
\end{equation*}%
imposed on the boundary $\partial \Omega $\ of a unit cell $\Omega $\ and by
computing the induced hyperstress components via%
\begin{equation*}
T_{ijk}=\frac{1}{2}<(\sigma _{ij}x_{k}+\sigma _{ik}x_{j})>.
\end{equation*}%
In this expression $<\cdot >$\ is the average operator defined by%
\begin{equation*}
<\cdot >=\frac{\overline{\rho }}{\left\vert \Omega \right\vert }\int_{\Omega
}\cdot \text{ }\mathrm{dV}
\end{equation*}%
where $\left\vert \Omega \right\vert $\ denotes the apparent area of $\Omega 
$\ and the relative density $\overline{\rho }$\ allows the correction of the
usual average operator due to the presence of the voids in a unit cell (\cite%
{ZMK09}).\ The components $T_{111}$, $T_{122}$ and $T_{212}$ are all
non-zero for any of the three cells in question. At the same time, the
components $T_{211}$ and $T_{121}$ are not null only when a chiral unit is
concerned. To see the last point, we write the explicit expressions allowing
the computation of $T_{211}$\ and $T_{121}$:%
\begin{equation*}
T_{211}=\frac{1}{2}<(\sigma _{12}x_{1}+\sigma _{12}x_{1})>=<\sigma
_{12}x_{1}>,\quad T_{121}=\frac{1}{2}<(\sigma _{12}x_{1}+\sigma _{11}x_{2})>
\end{equation*}%
where the moments of $\sigma _{12}x_{1}$, and $\sigma _{11}x_{2}$ are
involved. By examining the relevant fields obtained through FEM, it is seen
that: (i) for the $\mathrm{D}_{4}$-invariant cell, the field $\sigma _{12}$
is symmetric with respect to the middle vertical line (Fig.21), so that $%
\sigma _{12}x_{1}$ is well equilibrated (Fig.22) and the area average of $%
\sigma _{12}x_{1}$ is null; (ii) for a chiral cell, for example, an
orthotropic levogyre cell, the field $\sigma _{12}$ has not the symmetry
relative to the middle vertical line (Fig.23) and the area average of the
field $\sigma _{12}x_{1}$ is no more null. In this sense, we can say that
the non-zero chiral components of $\mathbb{A}$ come from the lack of a
reflection symmetry of the chiral unit cell in question. 
\begin{figure}[tbp]
\centering
\includegraphics[scale=0.4]{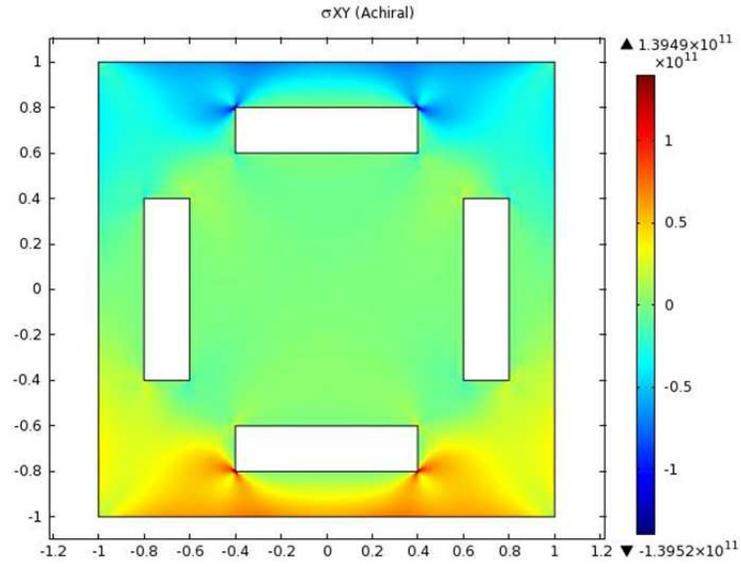}
\caption{The stress field $\protect\sigma _{12}$ in the cell which is not
chiral.}
\label{S12_D4}
\end{figure}
\begin{figure}[tbp]
\centering
\includegraphics[scale=0.4]{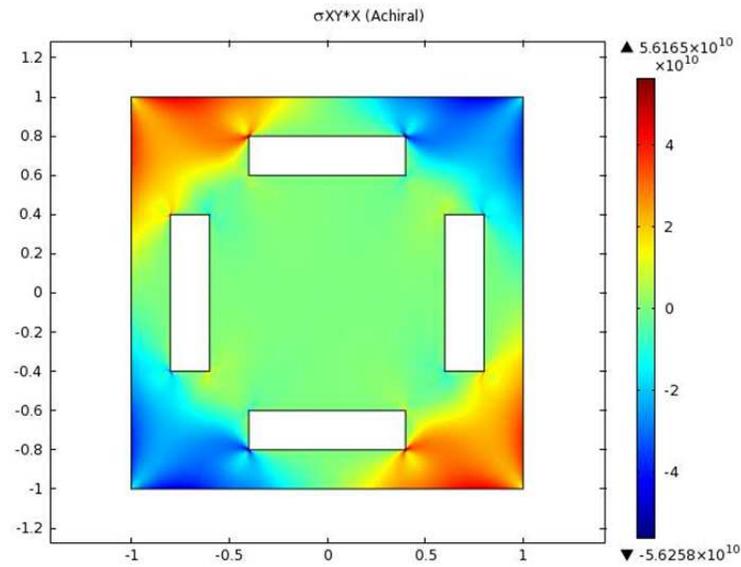}
\caption{The stress moment field $\protect\sigma _{12}x_{1}$ in the cell
which is not chiral. The area average of this field is null.}
\label{S121_D4}
\end{figure}
\begin{figure}[tbp]
\centering
\includegraphics[scale=0.4]{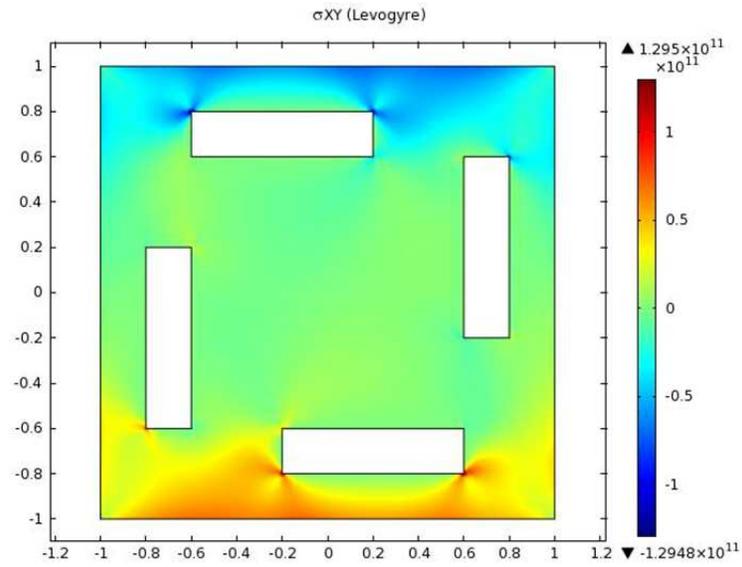}
\caption{The field $\protect\sigma _{12}$ in the levogyre chiral cell.}
\label{S12_Z4}
\end{figure}
\begin{figure}[tbp]
\centering
\includegraphics[scale=0.4]{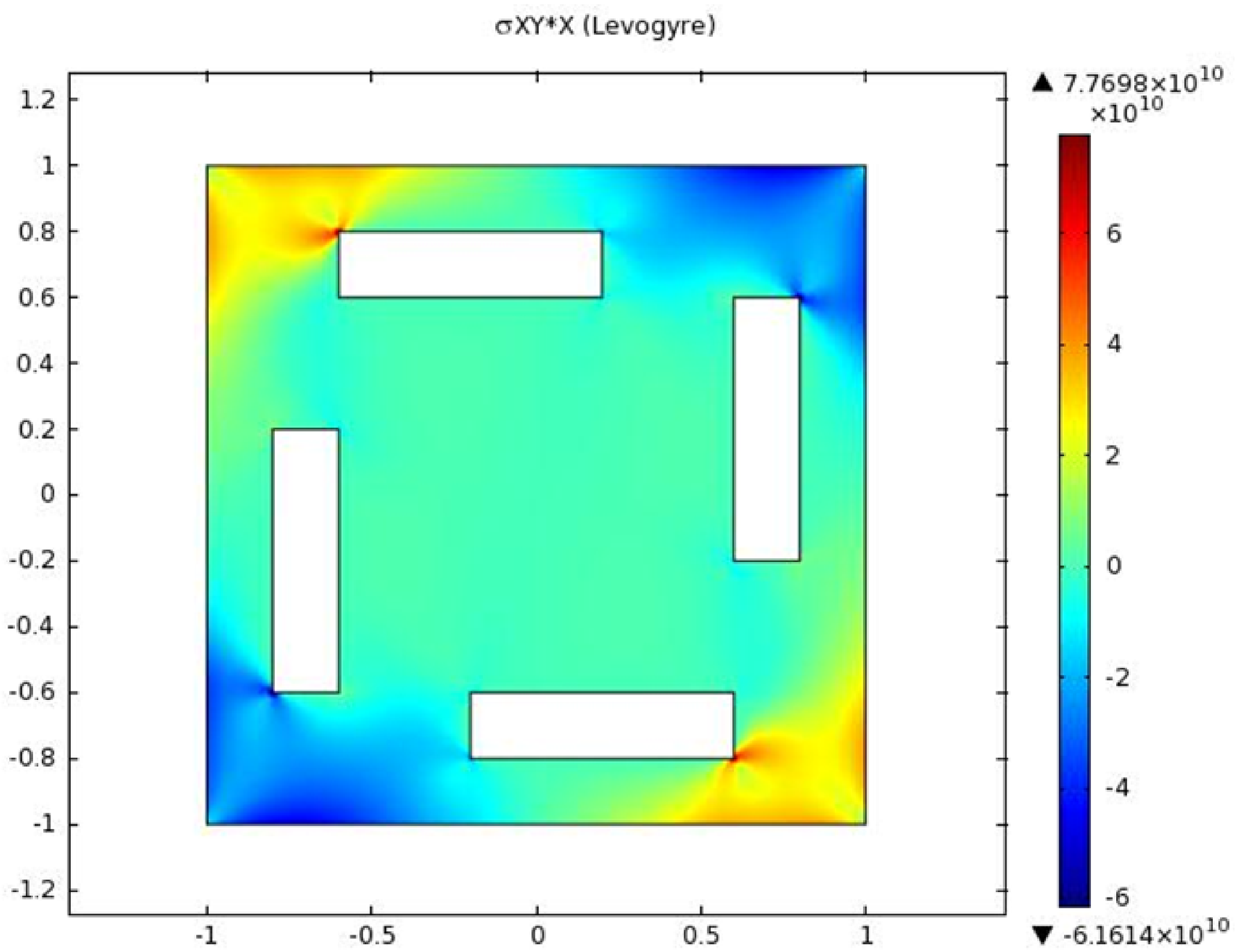}
\caption{The stess moment field $\protect\sigma _{12}x_{1}$ in the levogyre
chiral cell. The area average of this field is not null.}
\label{S121_Z4}
\end{figure}

\section{Concluding remarks}

Up to now, the development and application of (first) strain gradient
elasticity (SGE) have been almost exclusively confined to the isotropic
case. The complexity and richness of anisotropic SGE remain scarcely
exploited. In the present work and the companion one (\cite{LAH+12}), we
have studied materials whose microstructure exhibits centrosymmetry. For
these materials, SGE is defined by a sixth-order elastic tensor $\mathbb{A}$
in addition to the conventional fourth-order elastic tensor $\mathbb{C}$. In
the companion work (\cite{LAH+12}), the tensor $\mathbb{A}$ has been shown
to have 16 anisotropic symmetry classes apart from the isotropic symmetry
class. In the present work, the explicit matrix representations of $\mathbb{A%
}$ have been presented for all the anisotropic symmetry classes and written
in a compact and well-structured way. These results will be with no doubt
useful for the experimental identification, theoretical investigation and
numerical implementation of anisotropic SGE.

In the general case where the microstructure of a material does not exhibit
centrosymmetry, a fifth-order tensor $\mathbb{M}$ intervenes in addition to $%
\mathbb{C}$ and $\mathbb{A}$. Apart from a paper of \cite{Pap11} in which
the author studied $\mathbb{M}$ in the $\mathrm{SO}(3)$ symmetry class, the
questions concerning the symmetry classes and complete matrix
representations of $\mathbb{M}$ are still entirely open. A natural
continuation of the present work and the companion one (\cite{LAH+12}) will
consist in finding appropriates answers to these questions relative to $%
\mathbb{M}$.

Finally, we remark that, even though the theory of SGE was proposed about a
half century ago, its development and applications in anisotropic cases are
still at their beginning. This is particularly pronounced when the
fifth-order elastic tensor $\mathbb{M}$ is involved.


\begin{thebibliography}{Marangantia and Sharma(2007)}
\bibitem[Alibert et al.(2003)]{ASD03} Alibert, J.-J., Seppecher, P.,
dell'Isola, F., 2003. Truss modular beams with deformation energy depending
on higher displacement gradients. Math. Mech. Solids, 8, 51-73.

\bibitem[Auffray et al.(2009a)]{ABB09} Auffray, N., Bouchet, R., Bréchet,Y.,
2009. Derivation of anisotropic matrix for bi-dimensional strain-gradient
elasticity. Int. J. Solids Struct., 46, 440-454.

\bibitem[Auffray et al.(2010)]{ABB10} Auffray N., Bouchet R., Bréchet Y.,
2010. Strain-gradient elastic homogenization of a bidimensional cellular
material. Int. J. Solids Struct., 47, 1668-1710.

\bibitem[B\'ona et al.(2004)]{BBS04} Bona, A., Bucataru, I., Slawinski, M.A,
2004. Characterization of elasticity-tensor symmetries using $\mathrm{SU}(2)$%
. J. Elast., 75, 267-289.

\bibitem[B\'ona et al.(2007)]{BBS07} Bona, A., Bucataru, I., Slawinski, M.A,
2007. Material symmetries versus wavefront symmetries. Quarterly Jnl. of
Mechanics \& App. Maths., 60, 73-84.

\bibitem[Chadwick et al.(2001)]{CVC01} Chadwick, P., Vianello, M., Cowin, S.
C., 2001. A new proof that the number of linear elastic symmetries is eight.
J. Mech. Phys. Solids, 49, 2471-2492.

\bibitem[dell'Isola et al.(2009)]{DSV09} dell'Isola, F., Sciarra, G.,
Vidoli, S., 2009. Generalized Hooke's law for isotropic second gradient
materials. Proc. R. Soc. A, 465, 2177-2196.

\bibitem[dell'Isola et al.(2011)]{DMP11} dell'Isola, F., Madeo, A., Placidi,
L., 2011. Linear plane wave propagation and normal transmission and
reflection at discontinuity surfaces in second gradient 3D Continua. Z.
Angew. Math. Mech., 92, 52-71.

\bibitem[dell'Isola et al.(2012)]{DSM12} dell'Isola, F., Seppecher, P.,
Madeo, A., 2012. How contact interactions may depend on the shape of Cauchy
cuts in N-th gradient continua: approach "à la D'Alembert". Z. Angew. Math.
Phys., 92, 1-23.

\bibitem[Eringen(1968)]{Eri68} Eringen, A.C., 1968. Theory of micropolar
elasticity, in: Leibowitz, H. (Ed.), Fracture, vol. 2. Academic Press, New
York, pp. 621-629.

\bibitem[Fleck and Hutchinson(1997)]{FH97} Fleck, N.A., Hutchinson, J.W.,
1997. Strain gradient plasticity, in: Hutchinson, J.W., Wu, T.Y. (Eds.),
Advances in Applied Mechanics, Vol.33. Academic Press, New York, pp. 295-361.

\bibitem[Forest(1998)]{For98} Forest S., 1998. Mechanics of Generalized
Continua: Construction by Homogenization. J. Phys. IV, 8 39-48.

\bibitem[Forest and Trinh (2011)]{FT11} Forest, S., Trinh, D.K., 2011.
Generalized continua and non-homogeneous boundary conditions in
homogenization methods. Z. Angew. Math. Mech., 91, 90-109.

\bibitem[Forte and Vianello(1996)]{FV96} Forte, S., Vianello, M., 1996.
Symmetry classes for elasticity tensors. J. Elast., 43, 81-108.

\bibitem[Forte and Vianello(1997)]{FV97} Forte, S., Vianello, M., 1997.
Symmetry classes and harmonic decomposition for photoelasticity tensors.
Int. J. Eng. Sci., 35, 1317-1326.

\bibitem[Gologanu et al.(1997)]{GLP+97} Gologanu, M., Leblond, J.-B.,
Perrin, G., Devaux, J., 1997. Recent extensions of Gurson's model for porous
ductile metals, in Suquet, P.  (Ed.), Continuum Micromechanics, Springer-Verlag, New-York, pp. 61-130.
  
\bibitem[Gratias et al.(2000)]{GQK00} Gratias, D., Quinquandon, M., Katz,
A., 2000. Introduction to icosahedral quasicrystals. World Scientific Publ.

\bibitem[He and Zheng(1996)]{HZ96} He, Q.C., Zheng, Q.S., 1996. On the
symmetries of 2D elastic and hyperelastic tensors. J. Elast., 43, 203-225.

\bibitem[Huo and Del Piero(1991)]{HD91} Huo, Y.Z., Del Piero, G., 1991. On
the completeness of the crystallographic symmetries in the description of
the symmetries of the elasticity tensor. J. Elast., 25, 203-246.

\bibitem[Kruch and Forest(1998)]{KF98} Kruch, S., Forest, S., 1998.
Computation of coarse grain structures using a homogeneous equivalent
medium. J. Phys. IV,, 8, 197-205.

\bibitem[Koiter(1964)]{Koi64} Koiter, W. T., 1964. Couple-stresses in the
theory of elasticity: I and II. P. K. Ned. Akad. A. Math. 67, 17-44.

\bibitem[Kouznetsova et al.(2004)]{KGB04} Kouznetsova, V.G., Geers, M.G.D.,
Brekelmans, W.A.M., 2004. Multi-scale second-order computational
homogenization of multi-phase materials: a nested finite element solution
strategy. Comput. Method. Appl. M., 193, 5525-5550.

\bibitem[Lam et al.(2003)]{LYC+03} Lam, D.C.C., Yang, F., Chonga, A.C.M.,
Wang, J., Tong, P., 2003. Experiments and theory in strain gradient
elasticity. J. Mech. Phys. Solids, 51, 1477-1508.

\bibitem[Le Quang et al.(2012)]{LAH+12} Le Quang, H., Auffray, N., He,
Q.-C.,Bonnet, G., 2012. Symmetry groups and classes of sixth-order
strain-gradient elastic tensors tensors. (submitted).

\bibitem[Liu et al.(2011)]{LHH12} Liu, X.N., Huang, G.L., Hu, G.K. 2012.
Chiral effect in plane isotropic micropolar elasticity and its application
to chiral lattices. J. Mech. Phys. Solids, 60, 1907-1921.

\bibitem[Love(1944)]{Lov44} Love, A.E.H., 1944. Mathematical Theory of
Elasticity, Dover, New York.

\bibitem[Marangantia and Sharma(2007)]{MS07} Marangantia, R., Sharma, P.,
2007. A novel atomistic approach to determine strain-gradient elasticity
constants: Tabulation and comparison for various metals, semiconductors,
silica, polymers and the (Ir)relevance for nanotechnologies. J. Mech. Phys.
Solids, 55, 1823-1852.

\bibitem[Mindlin and Eshel(1968)]{ME68} Mindlin, R.D., Eshel, N. N., 1968.
On first strain-gradient theories in linear elasticity. Int. J. Solids
Struct., 4, 109-124.

\bibitem[Mindlin(1964)]{Min64} Mindlin, R.D., 1964. Micro-structure in
Linear Elasticity. Arch. Ration. Mech. An., 16, 51-78.

\bibitem[Mindlin(1965)]{Min65} Mindlin, R.D., 1965. Second gradient of
strain and surface-tension in linear elasticity. Int. J. Solids Struct., 1,
417-438.

\bibitem[Moakher and Norris(2006)]{MN06} Moakher, M., Norris A., 2006. The
closest elastic tensor of arbitrary symmetry to an elasticity tensor of
lower symmetry. J. Elast., 85, 215-263.

\bibitem[Nix and Gao(1998)]{NG98} Nix, W.D., Gao, H., 1998. Indentation size
effects in crystalline materials: a law for strain gradient plasticity. J.
Mech. Phys. Solids, 46, 411-425.

\bibitem[Pau and Trovalusci(2012)]{PT12} Pau, A., Trovalusci, P., 2012.
Block masonry as equivalent micropolar continua: the role of relative
rotations. Acta. Mech., 223, 1455-1471.

\bibitem[Papanicolopulos(2011)]{Pap11} Papanicolopulos, S.-A., 2011.
Chirality in isotropic linear gradient elasticity. Int. J. Solids Struct.,
48, 745-752.

\bibitem[Penrose(1974)]{Pen74} Penrose, R., 1974. Role of aesthetics in pure
and applied research. Bull. Inst. Math. Appl, 10, 266-271.

\bibitem[Sternberg(1995)]{Ste95} Sternberg, S., 1995. Group Theory and
Physics. Cambridge University Press, Cambridge.

\bibitem[Seppecher et al.(2011)]{SAD11} Seppecher, P., Alibert, J.-J.,
dell'Isola, F., 2011. Linear elastic trusses leading to continua with exotic
mechanical interactions., J. Phys. Conf. Ser. 319.

\bibitem[Tekoglu and Onck(2008)]{TO08} Tekoglu, C., Onck, P.R., 2008. Size
effects in two-dimensional Voronoi foams: A comparison between generalized
continua and discrete models. J. Mech. Phys. Solids, 56, 3541-3564.

\bibitem[Toupin(1962)]{Tou62} Toupin, R.A., 1962. Elastic materials with
couple stresses. Arch. Ration. Mech. An., 11, 385-414.

\bibitem[Truesdell and Toupin(1960)]{TT60} Truesdell, C., Toupin, R., 1960.
The classical field theories, in: Flügge, S. (Ed.), Handbuch der Physik Vol.
III/l, Springer-Verlag, Berlin, pp 226-793.

\bibitem[Truesdell and Noll(1965)]{TN65} Truesdell, C., Noll, W., 1965. The
Nonlinear Field Theories of Mechanics, in: Flügge, S. (Ed.), Handbuch der
Physik Vol. III/3, Springer-Verlag, Berlin.

\bibitem[Trinh and Forest (2010)]{TF10} Trinh, D.K., Forest, S., 2010. The
role of the fluctuation field in higher order homogenization. PMM-J. Appl.
Math. Mec., 10, 431-432.

\bibitem[Trinh et al.(2012)]{TJA+12} Trinh, D. K., Jänicke, R., Auffray, N.,
Diebels, S., Forest, S., 2012. Evaluation of generalized continuum
substitution models for heterogeneous materials. Int. J. Multiscale Com. Eng., 10, 527-549.

\bibitem[Xiao(1997)]{Xia97} Xiao, H., 1997. On Isotropic Invariants of the
Elasticity Tensor. J. Elast., 46, 115-149.

\bibitem[Zybell et al.(2009)]{ZMK09} Zybell, L., Mühlich, U., Kuna, M.,
2008. Constitutive equations for porous plane-strain gradient elasticity
obtained by homogenization. Arch. Appl. Mech., 79, 359-375.

\bibitem[Zheng and Boehler(1994)]{ZB94} Zheng, Q.-S., Boehler, J. P., 1994.
The description, classification, and reality of material and physical
symmetries. Acta. Mech., 102, 73-89.
\end{thebibliography}
\end{document}